\documentclass[aps,prl,reprint,a4paper,superscriptaddress,amsmath,amsfonts,amssymb, longbibliography,groupedaddress]{revtex4-1}

\usepackage{bm}
\usepackage[caption=false]{subfig}
\usepackage{tikz}
\usetikzlibrary{shapes.misc,positioning,shapes}
\usetikzlibrary{patterns}
\usepackage{array, multirow}
\usepackage[colorlinks,linkcolor=blue,citecolor=blue]{hyperref}%
\usepackage{relsize}

\usepackage{times}

\begin{document}
\preprint{APS/123-QED}
\title{Accelerated Magnonic Motional Cooling with Deep Reinforcement Learning}

\author{Bijita Sarma}
\email{bijita.sarma@oist.jp}
\author{Sangkha Borah}
\author{A Kani}
\author{Jason Twamley}
\affiliation{Quantum Machines Unit, Okinawa Institute of Science and Technology Graduate University, Okinawa 904-0495, Japan}

\date{\today}

%TC:ignore
%%%%%%%
\begin{abstract}
Achieving fast cooling of motional modes is a prerequisite for leveraging such bosonic quanta for high-speed quantum information processing. 
In this work, we address the aspect of reducing the time limit for cooling below that constrained by the conventional sideband cooling techniques; and propose a scheme to apply deep reinforcement learning (DRL) to achieve this. 
In particular, we have shown how the scheme can be used effectively to accelerate the dynamic motional cooling of a macroscopic magnonic sphere, and how it can be uniformly extended for more complex systems, for example, a tripartite opto-magno-mechanical system to obtain cooling of the motional mode below the time bound of coherent cooling. 
While conventional sideband cooling methods do not work beyond the well-known rotating wave approximation (RWA) regimes, our proposed DRL scheme can be applied uniformly to regimes operating within and beyond the RWA, and thus this offers a new and complete toolkit for rapid control and generation of macroscopic quantum states for application in quantum technologies.
\end{abstract}
%%%%%%%

%Use showkeys class option if keyword
%TC:endignore
\maketitle

%%%%%%%%%%%%%
\emph{Introduction.$-$}
Fast cooling of bosonic mechanical modes of macroscopic systems is a primary objective of the ongoing efforts in quantum technology~\cite{Schafermeier2016Nov,Guo2019Nov,Park2009Jul,Clark2017Jan,Frimmer2016Oct}, and is a prerequisite for diverse prospective applications, such as the realization of macroscopic superposition states~\cite{abdi2016dissipative}, gravitational tests of decoherence~\cite{Bose1999May,Marshall2003Sep}, ultra-precise measurements and sensing~\cite{Schliesser2009Jul,Manley2021Feb}, and bosonic quantum computing~\cite{bourassa2021fast}.
Macroscopic yttrium-iron-garnet (YIG, $\rm{Y}_3\rm{Fe}_5\rm{O}_{12}$) magnets have recently attracted strong interest towards such applications given the versatility of such systems in coupling to other modes falling in a wide frequency spectrum, e.g., with optical, microwave, and acoustic modes, as well as to superconducting qubits~\cite{zhang2016cavity,lachance2019hybrid,wang2020dissipative,li2018magnon,Zhang2014Oct,wang2018bistability}. 
In addition, highly polished YIG spheres feature high magnonic $Q$-factor and exhibit large frequency tunability properties due to the magnetic field dependence of the excited magnon modes. 
Considering, in particular, the cooling of motional modes of such objects, the usual method of sideband cooling based on weak magnomechanical interaction, operates on a time scale longer than the mechanical period of oscillation and depends on the relaxation dynamics of the subsystems.
Cooling of bosonic modes in such systems in a timescale less than the mode frequency is highly advantageous for quantum computation and bosonic error correction~\cite{joshi2021quantum,bourassa2021fast}. 
By going over to the strong coupling regimes in magnon-phonon interactions, the speed of motional cooling can be highly enhanced giving rise to accelerated cooling. 
However, in such strong coupling regimes, the energy-nonconserving dynamics prevails because of the simultaneous presence of counter-rotating interactions, which makes it impossible to use sideband motional cooling techniques in this regime. 
In this work, we explore the usefulness of a machine learning based approach to address the aspect of reducing the time limit for cooling below that constrained by the conventional cooling techniques.

Recently, various machine learning (ML) approaches, aided with artificial neural networks as function approximators, have found widespread technological applications~\cite{Goodfellow2016}. Among the various ML approaches, reinforcement learning (RL)~\cite{Sutton2018}, is considered to exhibit the closest resemblance to a human-like learning approach, in that the RL-agent tries to gather experience on its own by interacting with its environment in a trial and error approach. In RL terminology, the {\em environment} describes the virtual/real-world surrounding the agent, with all the physics hard-coded into it along with a reward function based on which the agent can classify its good moves from the bad ones. RL, when operated in combination with artificial neural networks, is known as deep reinforcement learning (DRL). DRL has become crucial in many industrial and engineering applications, primarily after recent seminal works by Google DeepMind researchers~\cite{silver_mastering_2016,silver_mastering_2017}. 
Following these developments, there have been several fascinating applications of DRL in various fundamental domains of science, including some in quantum physics in areas of quantum error correction,~\cite{Cirac2019, Mehta2018, Marquardt2018} quantum control~\cite{Borah2021}, and state engineering~\cite{Ueda2020,Niu2019Apr,Zhang2019Oct, Xu2021Apr, Wang2019a,Wrachtrup2020,Haug2020,Guo2021, Bilkis2020Aug,Prati2019}.

In this work, we propose a DRL-based dynamical coupling scheme for accelerated motional cooling of a macroscopic object, that works for a generalized parameter setting of the coupling strength between the subsystems. 
In particular, we use the protocol to cool the acoustic phonon modes of a YIG sphere with a magno-mechanical interaction, and show that it works efficiently in the strong coupling regime, where other methods such as sideband cooling fail.
Also, we show how going over to the strong coupling regime is particularly advantageous, as it lowers the cooling time well below the phonon oscillation period and two orders of magnitude below the sideband cooling time limit.
We demonstrate the usefulness and generalizability of our DRL cooling protocol by extending its application to a tripartite system of a trapped YIG magnet with its magnonic modes coupled to the center-of-mass (COM) mode in the trap and an optical cavity mode; and show that despite the system being in the ultrastrong coupling regime, our DRL scheme can reveal nontrivial coupling modulations to cool the motional mode, which is usually not possible with coherent counter-intuitive protocols.

%%%%
\begin{figure}[t]
    \centering
    \includegraphics[width=1.0\linewidth]{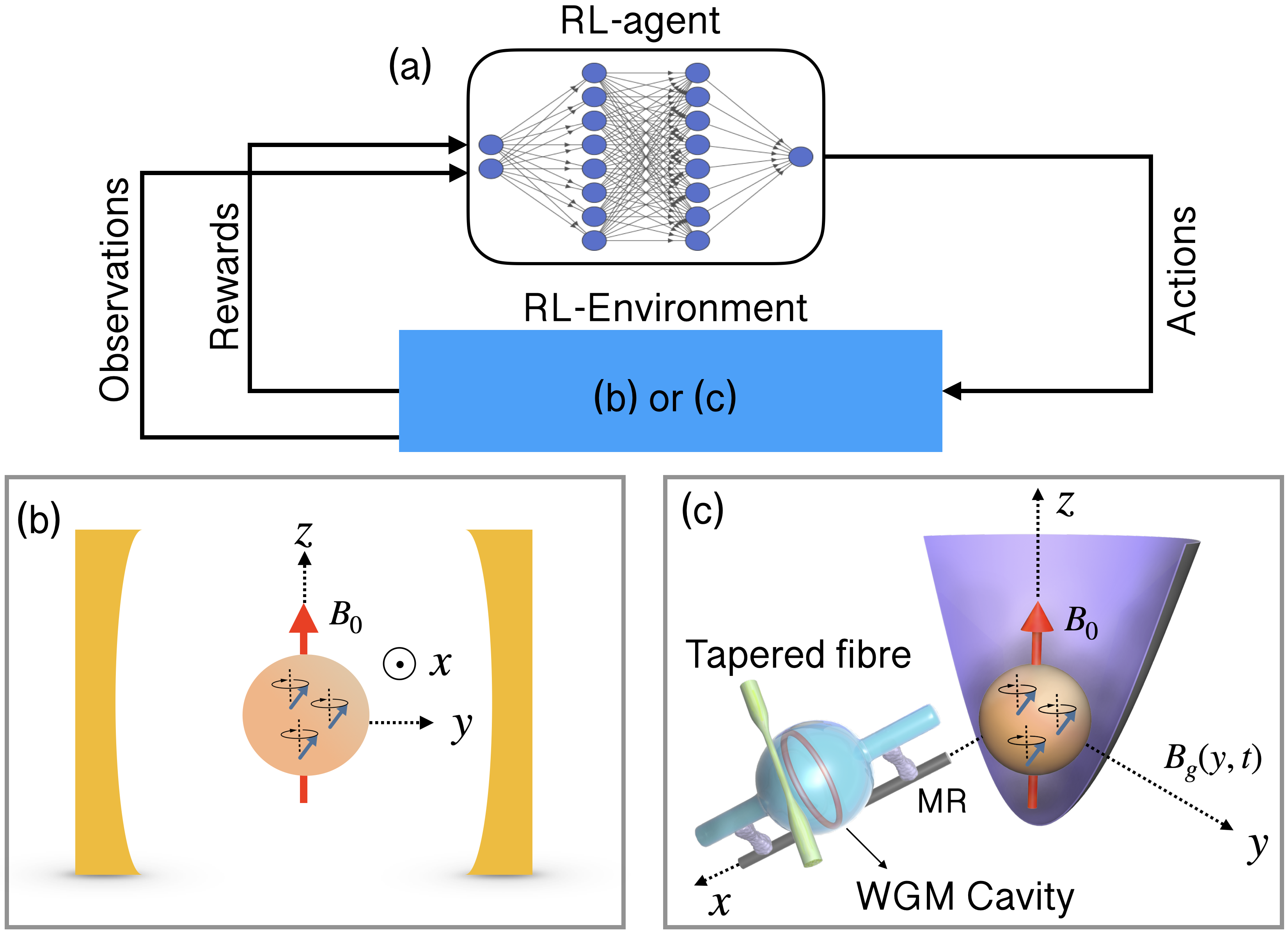}
    \caption{
    (a) The schematic workflow of the DRL protocol is shown, where the RL-environment is either the bipartite magno-mechanical system (b), or the tripartite opto-magno-mechanical system (c). See text for further detail on DRL and the physical models. 
    }
    \label{fig:model}
\end{figure}
%%%%%

%%%%%%%%%
\emph{Bipartite magno-mechanical cooling.$-$}
We consider a setup as shown in Fig.~\ref{fig:model}(b), where a highly polished YIG sphere is placed in a microwave cavity. 
With a large external homogeneous magnetic field, $B_{0}$, applied in the $\hat{z}$-direction, the YIG sphere is magnetized to its
saturation magnetization, $M_{s}=5.87 \times 10^{5} \mathrm{Am}^{-1}$, which gives rise to collective spin wave magnon modes~\cite{Tabuchi2014Aug,Zhang2014Oct}.
The frequency of the Kittel mode, which is the uniform magnon mode in the YIG sphere, is given by, $\omega_m = \gamma B_0$, where $\gamma/2\pi = 28 ~\mathrm{GHz/T}$ is the gyromagnetic ratio. 
Due to the magnetostriction properties, YIG spheres also exhibit high-Q acoustic modes which are coupled to the magnon modes~\cite{zhang2016cavity,li2018magnon}, and by driving the magnon modes with MW fields, the magno-mechanical coupling can be tuned and controlled~\cite{wang2018bistability,li2018magnon}. 
In the limit of adiabatic elimination for a low-$Q$ cavity, the bipartite magno-mechanical Hamiltonian is given by (see Supplemental Material for detail),
%%%
\begin{align}
\tilde{\mathcal{H}}/\hbar =  \Delta_m m^{\dagger} m + \omega_b b^{\dagger } b + (\tilde{G} m + \tilde{G}^\ast m^\dagger)(b + b^\dagger),
\label{eq:two_mode_hamiltonian}
\end{align}
%%%
where $m(m^\dagger)$ and $b(b^\dagger)$ are the magnonic and acoustic mode annihilation (creation) operators, $\omega_b$ is the resonance frequency of the acoustic mode and $\Delta_m=\omega_m - \omega_d$ (drive frequency $\omega_d$) is the detuning.
In such a bipartite system, while the beam-splitter interaction, $\tilde G m b^{\dagger } + \tilde G^* m^{\dagger} b$, valid for weak magno-mechanical coupling, favours mechanical cooling at the red sideband $\Delta_m = \omega_b$; the full coupling interaction accounting for the strong/ultrastrong coupling regimes, is not favourable in the usual sideband cooling approach. Sideband cooling works through anti-Stokes scattering of the excitation from the thermally populated mode, $b$ to the mode at zero entropy, $m$.
However, such cooling needs constant driving as it is a steady-state process that takes a duration of the order of the relaxation dynamics of the subsystems (see Supplemental Material). 
If one can access the strong coupling regime and manage to tame the counter-rotating interactions therein, there is a possibility of getting faster cooling than this limit. In the following, we design an algorithm based on DRL to model a dynamic variation of coupling, $\tilde{G}$, to get faster cooling of the acoustic mode, that operates within and beyond the weak coupling regime.

The schematic workflow of the DRL scheme applied to the physical model (RL-environment) is shown in Fig.~\ref{fig:model}(a). The RL-agent consists of a neural network model that is optimized for selected choices of actions that lead to desirable changes, and to net maximum rewards. The RL-agent is modelled using the recently proposed Soft Actor-Critic (SAC)~\cite{sac}
algorithm that is based on the maximization of the entropy, $\mathcal{H}(\pi_{\theta}(\cdot | s_t) )$ of the policy, $\pi_{\theta}$ as well as the long-term discounted cumulative return $r(s_t, a_t)$, i.e.,~$
\max \mathbb E \left[ \sum_{t} \gamma_t ( r(s_t, a_t) + \alpha \mathcal{H}(\pi_\theta(\cdot | s_t) ) )\right],
$
where, $\theta$ denotes the optimizable weights of the neural networks, $\gamma$ is the discount factor and $\alpha$ is the regularization coefficient that determines the stochasticity of the policy. The policy, $\pi_\theta$ sets the rules for the particular actions to be applied on the RL-environment (see Supplemental Material for further details).

The dynamics of the system is described by the quantum master equation (QME) for the density matrix $\rho$ with the Hamiltonian $\tilde{\mathcal{H}}$ as,
%%%%%
\begin{align}
\nonumber
{d \rho}(t) & = - \frac{i}{\hbar}\left[\tilde{\mathcal{H}}, \rho \right] {dt}\\
& + \sum_{j}^{}{\left[\kappa_j\left(\bar{n}_{c_j} +1\right){\mathcal{L}}[c_j]\rho+\kappa_j \bar{n}_{c_j}{\mathcal{L}}[c_j^\dagger]\rho \right]} {dt},
\end{align}
with dissipations and thermal fluctuations given by the Lindblad superoperators, ${\mathcal{L}}[c_j]\rho\equiv c_j\rho c_j^\dagger - \frac{1}{2}\,\{c_j^\dagger c_j,\rho\}$ where, $\kappa_j$'s are the damping rates of the modes; and the thermal occupation of each bosonic mode is given by, $\bar{n}_{c_j}=[{\rm exp}(\hbar {\omega_j}/k_B T)-1]^{-1}$, where $T$ is the bath temperature and $k_B$ is the Boltzmann constant. Solving the full QME to obtain the mean occupancy is a computationally intensive task.
DRL typically requires several thousands of episodes of training, and solving the full QME within each episode is too resource sensitive for complex systems such as the ones we consider in this work. 
We employ an alternative approach to compute the mean occupancies in each mode using a set of linear differential equations for the second-order moments obtained from the QME, given by, $ {\partial }_t\left\langle {\hat{o}}_i{\hat{o}}_j\right\rangle =\mathrm{Tr}\left(\dot{\rho }{\hat{o}}_i{\hat{o}}_j\right)=\sum_{m,n}{{\zeta}_{mn}\left\langle {\hat{o}}_m{\hat{o}}_n\right\rangle }, $
where ${\hat{o}}_i$,\  ${\hat{o}}_j$,\  ${\hat{o}}_m$,\  ${\hat{o}}_n$ are one of the operators (${c_j}^{\dagger },\  c_j$); and ${\zeta }_{mn}$ are the corresponding coefficients. 
Instantaneous solutions of these equations and the controls, $\tilde G$, are used as the observations for the RL-agent in Fig.~\ref{fig:model}(a), and the reward function is chosen as, $r(t)=1/{\tilde{n}_b}(t)$.
Here ${\tilde{n}_b}(t) = \langle b^\dagger b \rangle (t)/n_b^{T}$ is the cooling quotient of the phonon number with respect to thermal occupancy, $n_b^{T}$  at temperature $T$, where $\langle b^\dagger b \rangle$ represents the mean value of the phonon population. Further details of the DRL controller can be found in Supplementary Information.

%%%%%%%
\begin{figure}[t]
    \centering
    \includegraphics[trim={0.0cm 0.0cm 0cm 0cm},clip,width=1.0\linewidth]{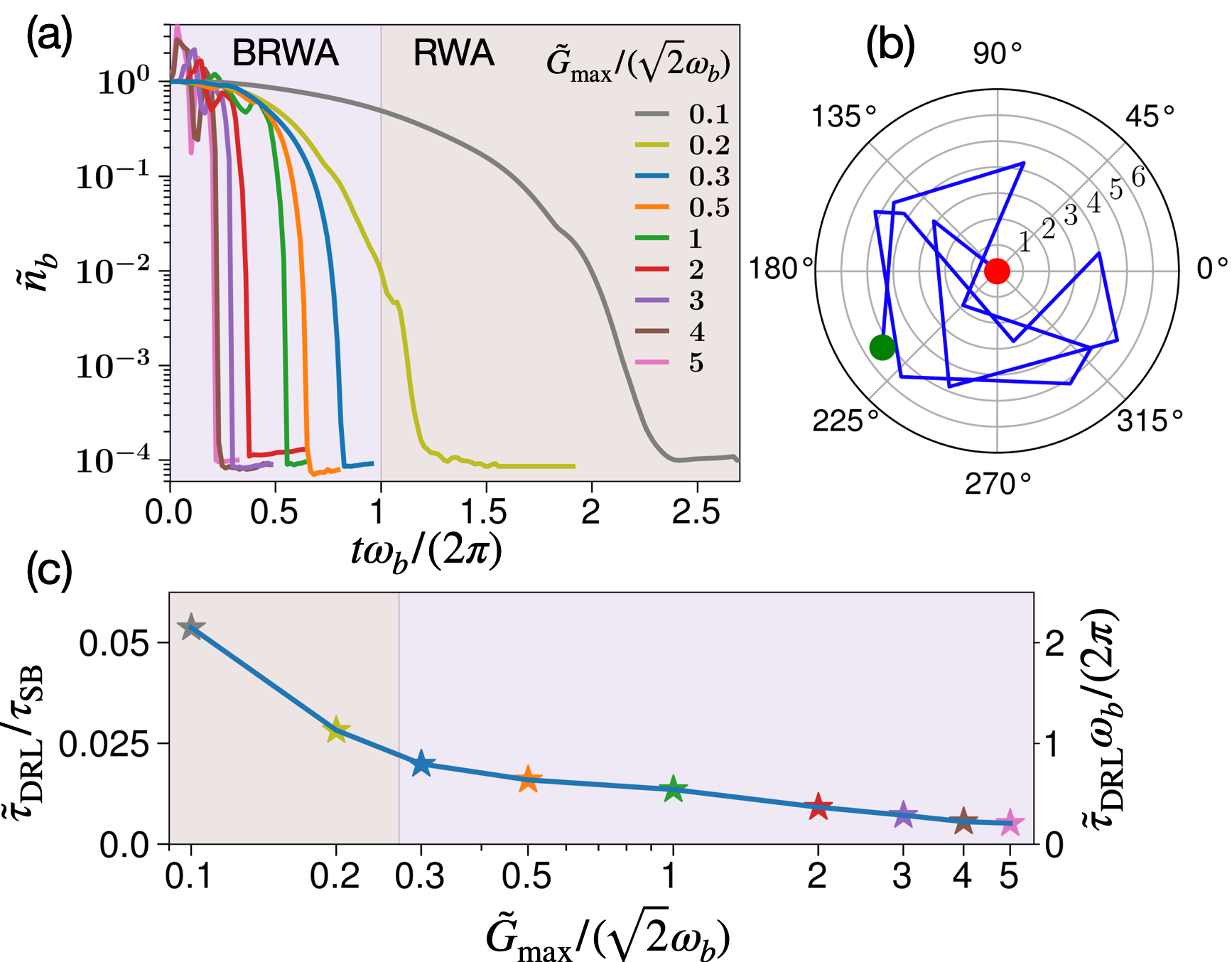}
    \caption{(a) Accelerated DRL-based cooling of the phonon mode shown with the cooling quotient, $\tilde n_b$, of the bipartite system as shown in Fig.~\ref{fig:model}(b) as a function of $\tilde{G}_{\mathrm{max}}$, the maximum allowed control [with the values under RWA or beyond-RWA (BRWA)]. 
    (b) The DRL-optimized complex pulse sequence is shown as a polar plot for $\tilde{G}_{\rm max}/(\sqrt{2}\omega_b) = 5$. The green dot shows the starting point while the red dot shows the final point.
    (c) The time required for cooling with the DRL-scheme, $\tilde{\tau}_{\rm DRL}$ (right $y$-axis), for $\tilde n_b \lesssim 10^{-4}$, is compared with the time limit for effective sideband cooling, $\tau_{\rm SB}$ (left $y$-axis).  
    }
    \label{fig:two_mode}
\end{figure}
%%%%%%%

In Fig.~\ref{fig:two_mode}(a), 
we show the cooling quotient, 
$\tilde n_b$ for the DRL-optimized controls, as a function of $\tilde G_{\rm max}$ (the maximum of the coupling parameter). 
We consider $\Delta_m = \omega_b$, and damping rates as, $\kappa_b/\omega_b = 10^{-5}$ and $\kappa_m/\omega_b = 0.1$.
It is found that as the coupling is increased towards the ultrastrong coupling regime ($\tilde{G} \gtrsim \omega_b$), the cooling becomes much faster. 
The DRL-optimized complex pulse sequence is shown as a polar plot for $\tilde{G}_{\rm max}/(\sqrt{2}\omega_b) = 5$ in Fig.~\ref{fig:two_mode}(b). 
We denote the minimum time for cooling achieved by our method as $\tilde{\tau}_{\mathrm{DRL}}$. Fig.~\ref{fig:two_mode}(c) compares this time as a function of $\tilde{G}_{\mathrm{max}}$ with respect to the sideband cooling time limit, ${\tau}_{\rm SB}$,
which represents the shortest cooling time limit possible with these methods, that work only when the rotating wave approximation (RWA) is applicable (see Supplemetal Material). 
With the DRL-based coupling modulations, we can achieve very low limits of cooling time compared to sideband cooling techniques, showing a lowering of approximately two orders of magnitude; which for the ultrastrong coupling regime, is further lowered.

%%%%%%%%%%
\emph{Tripartite opto-magno-mechanical cooling.$-$}
Now, we show that the proposed scheme can be extended effectively to more complex systems, for example a higher-order tripartite opto-magno-mechanical system, where we intend to cool the motional mode through non-trivial three-mode interactions. 
For this, we consider a system comprising a levitated YIG sphere in a harmonic trap~\cite{Seberson2020Dec, Rusconi2017quantum, Huillery2020Apr}, along with a driven WGM optical microresonator placed along the $\hat{x}$-direction with a magnetostrictive rod (MR) attached to it, as depicted in Fig.~\ref{fig:model}(c), which will be used as the RL-environment in Fig.~\ref{fig:model}(a) in the DRL model. 
In a large external homogeneous magnetic field, $B_{0}$, applied in the $\hat{z}$-direction, the YIG sphere is magnetized to the saturation magnetization, $M_{s}$, and the homogeneously magnetized fundamental magnon mode (Kittel mode) produces a change in the axial length of the MR, which modulates the WGM optical mode frequency~\cite{Forstner2014UltrasensitiveMagnetometry,Xia2015AnQubits, Yu2016OptomechanicalResonator,Guo2019Nov}. 
This gives rise to a coupling between the WGM optical mode $(a)$ and the magnon mode $(m)$ of the form, $\Omega_S a^\dagger a(m+m^\dagger)$, where $\Omega_S=\Delta \omega$ is the optical frequency shift. 
The magnon mode can also be coupled to the COM motion of the YIG sphere, by applying a spatially inhomogeneous external time-dependent magnetic field,
$\mathbf{H}_{g}(y, t)$~\cite{Hoang2016electron,Delord2018ramsey}, which satisfies the weak driving, $\left|\mathbf{H}_{g}(y, t)\right| \ll H_{0}$, and small-curl, $\left|\nabla \times \mathbf{H}_{g}(y, t) \| \mathbf{r}\right| \ll\left|\mathbf{H}_{g}(y, t)\right|$ conditions (see the Supplemental Material for more information).
Considering a time-varying gradient magnetic field of the form, $\mathbf{H}_g (\mathbf{y}, t)=\frac{b_g(t)}{\mu_{0}} y \hat{y}$, ($b_g$ in units of $[\mathrm{T} / \mathrm{m}]$), the interaction Hamiltonian for the COM motion in the $\hat{y}$-direction (frequency $\omega_b$) and the magnon mode is given by, 
$
\mathcal{H}_{mb} (t)= \tilde{\Omega}_P (\hat{b}+\hat{b}^{\dagger}) (\hat{m}+\hat{m}^{\dagger}), 
$
with $\tilde{\Omega}_P  =\frac{b_g}{4} \sqrt{\frac{|\gamma| M_{S}}{\rho \omega_{b}}}$, where $M_{s}=5.87 \times 10^{5} \mathrm{Am}^{-1}$ is the saturation magnetization, and $\rho = 5170\ \rm{kg/m^3}$ is the mass density of YIG. In the rotating frame of the cavity drive and the displacement picture of the average field in each mode~\cite{khosla2017quantum}, the complete Hamiltonian is described by
\begin{align}
\nonumber
\tilde{{\mathcal{H}}}/\hbar = & \Delta_{a}  a^{\dagger } a+\omega_{m} m^{\dagger} m+\omega
_{b} b^{\dagger } b  + \tilde{\Omega}_S(a + a^{\dagger})(m + m^{\dagger }) \\
+& ~ \tilde{\Omega}_P(m + m^{\dagger})(b + b^{\dagger }),  
\end{align}%
where ${\Delta}_{a}$ is the cavity detuning, and $\tilde{\Omega}_S$ is the driving-enhanced optomagnonic
coupling rate. One can modulate $\tilde{\Omega}_S$ via the external drive, whereas the phonon-magnon coupling $\tilde{\Omega}_P$ can be modulated using the time-varying magnetic field gradient. 
In order to cool the COM motion, we intend to transfer the phonon population from the COM mode to the optical mode without populating the magnon mode.
The damping rates of the cavity, magnon, and COM modes are given by, $\kappa_i$'s, and the corresponding thermal populations at temperature $T$ are $n_{i}^{T}$, with $i \in \{a, m, b\}$. Since the optical cavity mode oscillates at high frequency, its corresponding thermal bath even at room temperature yields zero thermal occupancy, however, the phonon and magnon baths are occupied.

Similar to the bipartite system discussed above, we next use the second-order moment equations to solve the dynamics of the system and the DRL scheme is used to optimize the controls, $\tilde{\Omega}_{\rm P/S}$  by  maximizing the net reward signal per episode,
$r(t) = 1/{\tilde{n}_b} - \lambda \, {\langle m^\dagger m }\rangle(t)$, 
where $\lambda $ is a constant chosen such that the magnon mode does not get populated.  
%%%%
Given the fact that the COM mode frequencies of the YIG  are of the order of $\omega_b/2\pi\sim 10- 100$'s of ${\rm kHz}$, and the magnon frequency is $\omega_m/2\pi \sim 10 ~{\rm GHz}$, the ideal choice of $\omega_m$ for $\omega_b/2\pi = 100\  \rm kHz$ is $\omega_m = 10^5\,\omega_b$. 
With such a high frequency difference with the intermediate magnon mode at $\omega_m = 10^5\,\omega_b$, this constitutes a largely detuned system ($\omega_m\gg \sqrt{\tilde{\Omega}_P^2 + \tilde{\Omega}_S^2}$). In such a system while the magnon mode is usually decoupled, the ideal time limit to obtain swap between mechanical quanta and optical mode with the ideal Raman pulses is given by $\tilde{\tau}_{\rm lim} = \pi \omega_m/(2\tilde{\Omega}_P \tilde{\Omega}_S)$ (see the Supplemental Material). 
However, this is the limit for the situation as long as the RWA is valid. 
It is also noted that the effectiveness of this kind of cooling is highly reduced in presence of damping, and going beyond the RWA to decrease the cooling time limit is not possible because of the counter-rotating dynamics.
We apply the DRL strategy to work in a regime where $\tilde{\Omega}_i$'s are sufficiently high to access the cooling limit not obtainable by these conventional means, and also keep the counter-rotating dynamics in control.
While the use of the method of coupled second order moments reduces the computation resources drastically, 
simulating the dynamics with the choice of the realistic parameter $\omega_m = 10^5\ \omega_b$ with high coupling strengths, for example $\tilde{\Omega}_S^{\rm max} = \tilde{\Omega}_P^{\rm max}=100\ \omega_b$,
turns out to be a computationally highly intensive problem, due to the very stiff solution of the set of differential equations. 
Hence, we adopt a two-step training procedure for the problem. 
In this protocol, we first use an \emph{auxiliary} system with $\omega_m = 10^3 \omega_b$ and $\tilde{\Omega}_S^{\rm max} = \tilde{\Omega}_P^{\rm max}=10\ \omega_b$, for which solution of the set of equations can be obtained without much computational effort. The trained auxiliary model is then used as a supervisor/teacher for the actual system, that we call \emph{primary}, with $\omega_m = 10^5\omega_b$ and $\tilde{\Omega}_S^{\rm max} = \tilde{\Omega}_P^{\rm max}=100\ \omega_b$. Training the primary system for a few hundred episodes with periodic evaluation of the RL-agent yields the best trained model. In the literature of RL, such a scheme is known as \emph{imitation learning}, and is a feature of generalizability of RL-trained model.

\begin{figure}[t]
    \centering
    \includegraphics[trim={0.0cm 0.0cm 0cm 0.0cm},clip,width=1.0\linewidth]{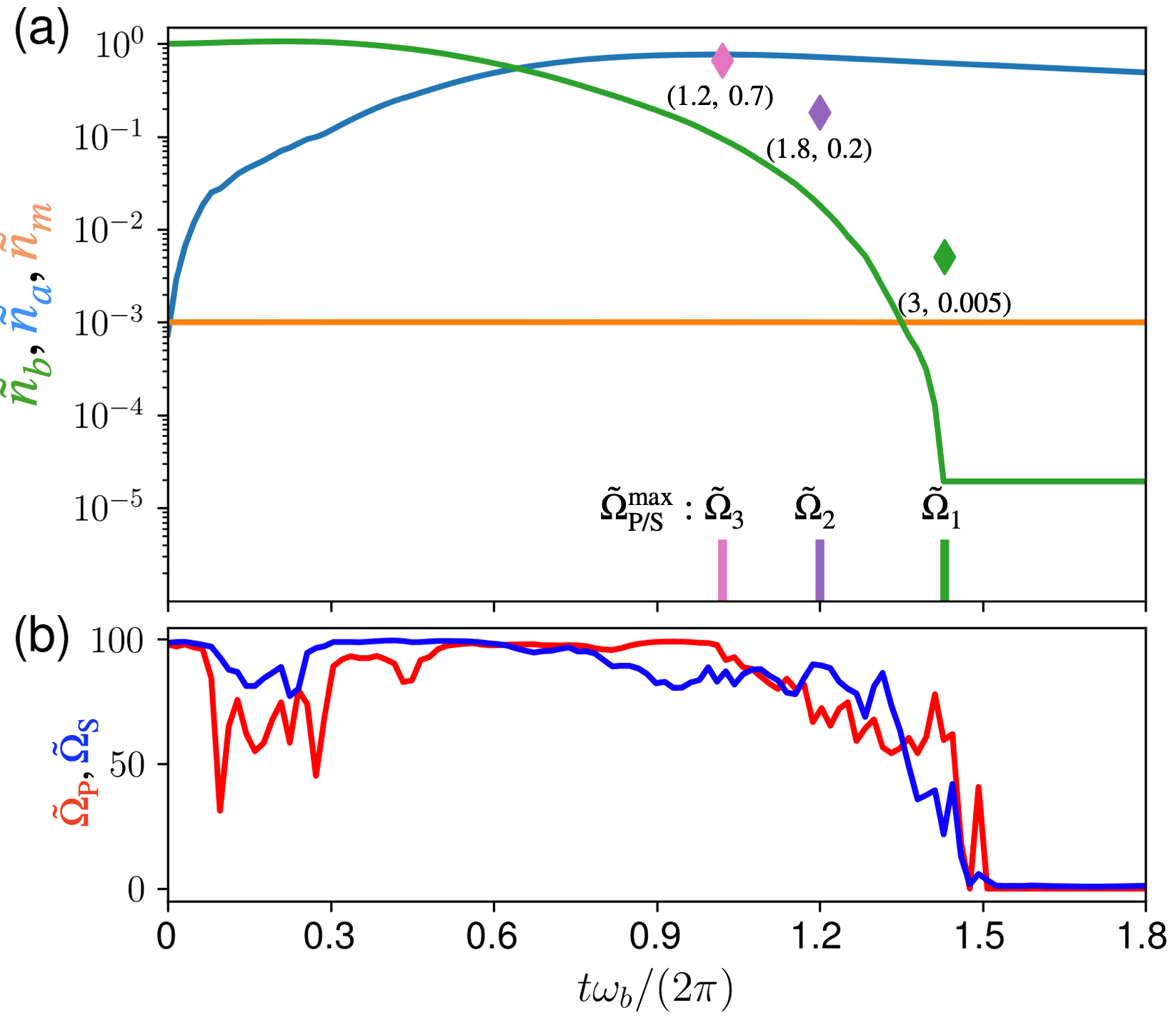}
    \caption{(a) The DRL-scheme based mean occupancies of the phonon, magnon and photon modes are shown in terms of the cooling quotient, $\tilde{n}_i$ (in green, orange and blue lines respectively) for $\tilde{\Omega}_{\{\rm S,\rm P\}}/\omega_b = 100$. The vertical lines show the time limits for the DRL-based cooling with $\tilde{\Omega}_i^{\rm{max}}/\omega_b=\{100,120,150\}$ respectively (in green, purple, pink). While the DRL-derived scheme gives a cooling quotient of $\tilde{n}_b<10^{-4}$, the corresponding Raman pulses with the same amount of maximum allowed coupling strength gives lower values, and the cooling time limits for these are far higher than those obtained with the DRL-scheme.
    The markers show the cooling results obtained for the conventional counter-intuitive pulses, with the corresponding cooling quotients and time limits shown with each point in the order $(\tilde{\tau}_{\rm{lim}},\tilde{n}_b)$. 
    (b) The corresponding DRL-optimized coupling parameters are shown for the case of maximum allowed value of $\tilde{\Omega}_{\{\rm S,\rm P\}}/\omega_b = 100$.}
    \label{fig:three_mode}
\end{figure}

In Fig.~\ref{fig:three_mode}(a), the cooling quotients of the photon, magnon and phonon modes, $\tilde{n}_i = n_i/n_T$, are shown, where $n_i$'s are the mean occupancies and $n_T$ is the phonon thermal bath population. The cooling time limit for this system is given by $\tilde{\tau}_{\rm lim}\omega_b/(2\pi)\sim 5$ (see the Supplemental Material).
The plot shows the population dynamics in the three modes for the DRL-based coupling parameters with a maximum allowed value, $\tilde{\Omega}_i^{\rm max}/\omega_b = 100$. One can see that while the magnon mode is kept at a constant population, there is a scattering between the mechanical and optical mode that gives rise to five orders of cooling in the mechanical mode. This draws an analogy to the stimulated Raman adiabatic passage (STIRAP) techniques for three-mode systems. However, it is well-known that such adiabatic techniques require longer time and only works ideally as long as the counter-rotating terms are not present (see the Supplemental Material). On the contrary, the coupling parameters found by our DRL protocol are non-trivial, which are shown in Fig.~\ref{fig:three_mode}(b), and the overall time required for cooling is reduced below the adiabatic limit even for high values of coupling. In the bottom panel of Fig.~\ref{fig:three_mode}(a) we show the cooling time required by our DRL method for even larger values, and higher coupling parameters yield even lower time limits.

%%%%%%%%%
%%%%%%

%%%%%%%%
\emph{Conclusion.$-$}
To conclude, in this work we address the aspect of reducing the time required for cooling bosonic motional modes below the time limit accessible by well-known cooling methods with scattering techniques. 
While such conventional methods of cooling do not work in strong and ultrastrong coupling regimes where the RWA is not valid, we design a DRL-based algorithm that works within and beyond such regimes, and show that by accessing the ultrastrong coupling limit with DRL-designed pulses the cooling limit can be broken resulting in accelerated cooling. 
We further show how the protocol can be adapted for cooling in a tripartite system with an opto-magno-mechanical interaction, that represents more complexity for the DRL-control owing to the huge dimensionality in the Hilbert space; and find nontrivial three-mode interactions leading to accelerated cooling breaking the coherent cooling time limits.
Thus, this study outlines a comprehensive toolbox for application of DRL for fast and efficient quantum control in a magnonic system within and beyond RWA restrictions, which can be adapted to other quantum systems of interest. For future and ongoing efforts in quantum technology, this is expected to play a pivotal role, especially in conjunction with various laboratory experiments.
%%%%%%%%%

%TC:ignore
\section{acknowledgments}
\begin{acknowledgments}
The authors thank Okinawa Institute of Science and Technology (OIST) Graduate University for the super-computing facilities provided by the Scientific Computing and Data Analysis section of the Research Support Division, and for financial support. The authors are grateful for the help and support provided by Jeffery Prine from the Digital Content, Brand, and Design Section of the Communication and Public Relations Division at OIST.
\end{acknowledgments}

\newpage
\widetext

\author{Bijita Sarma}
\email{bijita.sarma@oist.jp}
\author{Sangkha Borah}
\author{A Kani}
\author{Jason Twamley}
\affiliation{Quantum Machines Unit, Okinawa Institute of Science and Technology Graduate University, Okinawa 904-0495, Japan}

\def\theequation{S\arabic{equation}}
\renewcommand{\thepage}{S\arabic{page}} 
\renewcommand{\thesection}{S\arabic{section}}  
\renewcommand{\thetable}{S\arabic{table}}  
\renewcommand{\thefigure}{S\arabic{figure}}
\setcounter{figure}{0}
\setcounter{table}{0}
\setcounter{section}{0}
\setcounter{subsection}{0}
\setcounter{page}{1}
\begin{center}
	\textbf{\large Supporting Information}
\end{center}
\maketitle

\section{Reinforcement learning and soft actor-critic algorithm }
\noindent Reinforcement learning (RL) has been one of the most exciting fields of machine learning (ML) following a few revolutionary works carried out during 2013-2016 by the DeepMind and Google~\cite{silver_mastering_2017, silver_mastering_2016}. Although these applications were demonstrated initially for different board games, researchers all around the world are now looking into its potential uses for different technological applications. In RL, a software agent (we will call it the RL-agent), makes some observations and applies some actions on an environment (we call it the RL-environment), that causes a change in the dynamics of the environment and, in return, the RL-agent receives some scalar values as rewards. The objective of the RL-agent is to maximize the total rewards obtained in a particular time range (we call it an episode). This approach of training is particularly different from typical training approaches where the ML-model is trained based on predefined data with labels (supervised learning), or with no labels (unsupervised learning). 

\begin{figure*}[!hbt]
    \centering
    \includegraphics[width=0.6\linewidth]{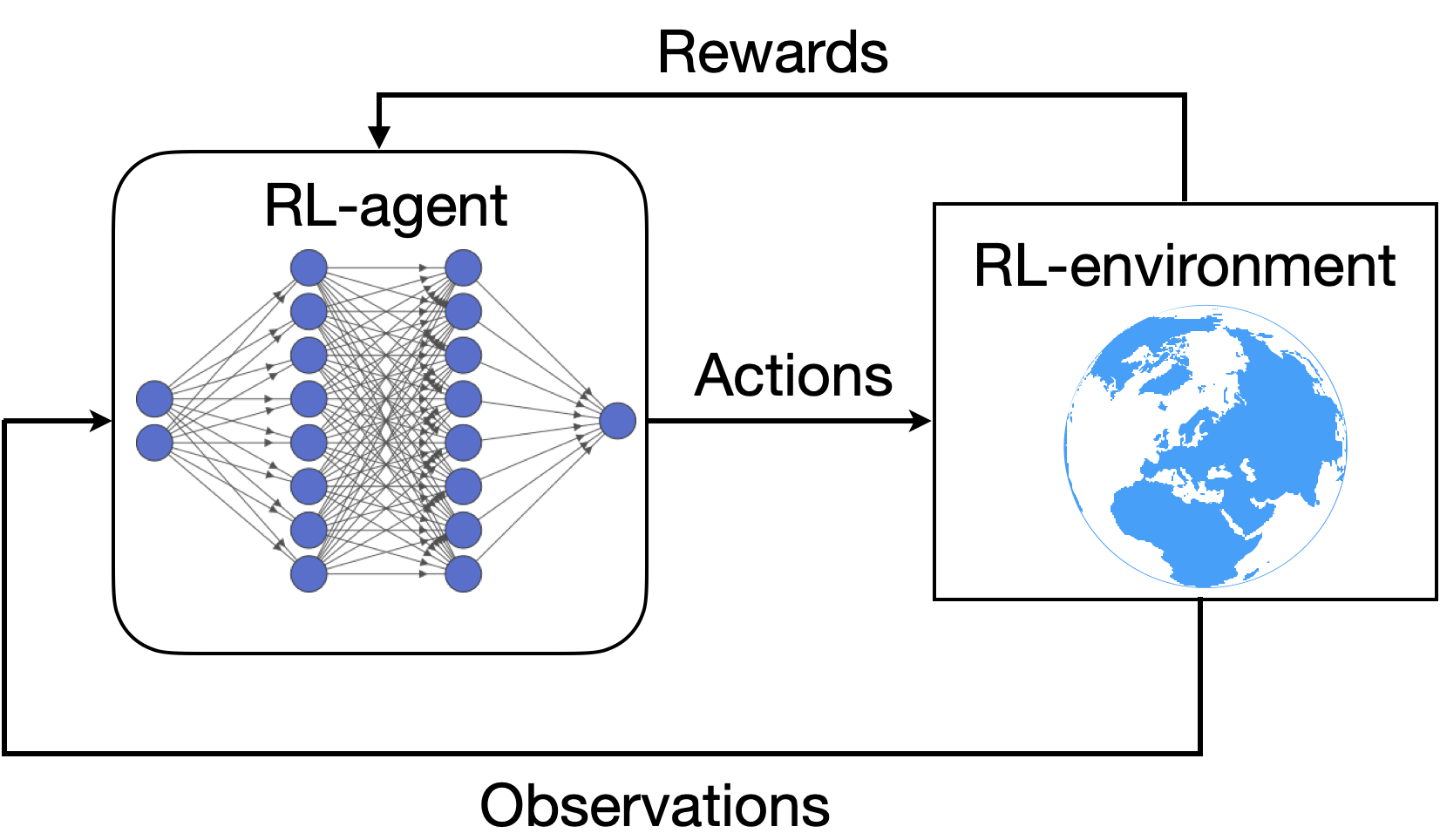}
    \caption{The workflow of deep reinforcement learning (DRL) is shown. The RL consists of two main characters -- the RL-environment and the RL-agent. The RL-agent, in the case of DRL, comprises artificial neural networks which operate as function approximators. The RL-agent applies certain actions on the RL-environment and receives certain quantities as observations along with a scalar reward signal, based on which the neural network is updated to adopt a better policy for improved rewards.   }
    \label{fig:DRL_workflow}
\end{figure*}

The workflow of a typical RL is shown in Fig.~\ref{fig:DRL_workflow}. Here, the block named as the RL-agent is  depicted as comprising a set of neural network layers which act as function approximators. This type of RL is known as deep reinforcement learning (or DRL for short). On the other hand, the block in the right depicts the RL-environment, which is the world wherein the RL-agent lives and can affect change via actions on the dynamics of the environment. In RL terminology, the environment signifies the real/virtual space in which the rules of the physics of the problem/system are encoded along with a reward estimation function returning some real scalar values that determine whether the applied action has led to desirable changes or not. In practice, the RL-agent makes a certain number of interactions by first randomly applying some actions on the RL-environment based on which the weights of the nonlinear neural network function estimators are updated. The procedure is repeated several hundreds/thousands of times, depending on the complexity of the problem. Such a learning process is analogous to the learning process of a human, where the agent learns by conducting trial and error experiments on the RL-environment. Depending on the type of the RL algorithm used, the neural network based agent can explore even newer possibilities not usually explored, thus outperforming the analogous supervised learning agent.

%%%%%%%%%%%%%%%%%%%%%%%%%%%%%%%%%
\begin{figure*}[!hbt]
    \centering
    \includegraphics[width=0.6\linewidth]{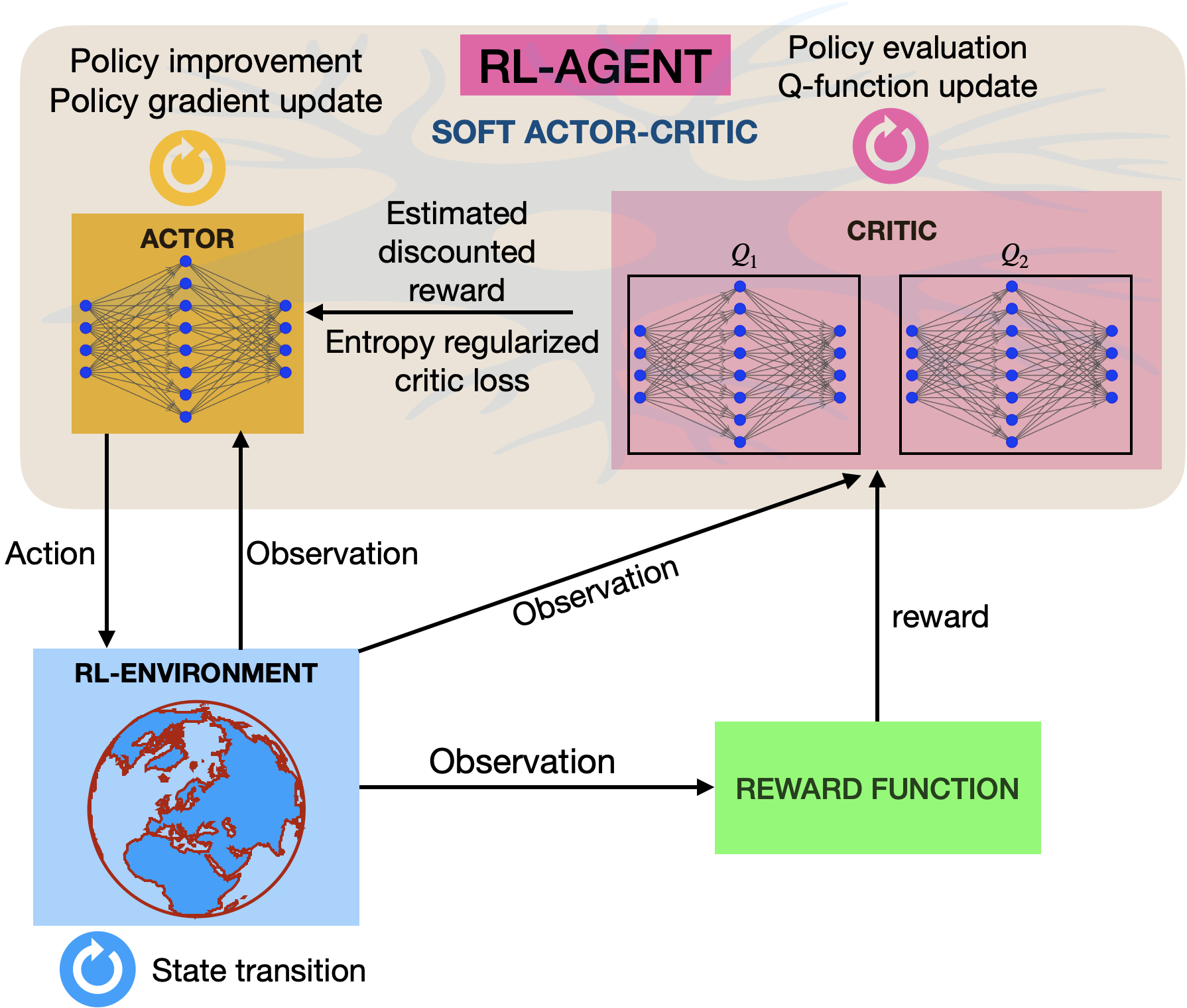}
    \caption{The workflow of DRL in the soft-actor-critic (SAC) setting is shown. In the actor-critic framework, the RL-agent has two independent neural networks for policy evaluation and policy improvement to guide the choice of actions, known respectively as the critic and actor networks. The critic, in the SAC framework, consists of two neural networks, $Q_1$ and $Q_2$ for improved policy evaluation through Q-function updates. Based on the suggestion of the critic (in terms of estimation of the discounted future rewards and entropy regularized critic loss for SAC), the actor-network is updated, and the next action is chosen. }
    \label{fig:rl_model}
\end{figure*}
%%%%%%%%%%%%%%%%%%%%%%%%%%%%%%%%%

The algorithm that the RL-agent uses to determine the actions is called its policy, which in the case of DRL, represents the neural network itself that takes the observations as the input, and outputs the actions. A policy can be deterministic in which case the action of the RL-agent is determined by the policy parameters $\theta$ for a given state $s_t$ at time $t$:  $a_t = \mu_\theta (s_t)$, or stochastic in which the actions are sampled from a probability distribution conditioned on $s_t$: $a_t  \sim \pi_\theta (\cdot |s_t)$. The task of the RL-agent is to optimize the policy parameters ($\theta$) to maximize the net discounted  rewards $R(\tau) = \sum_{t=0}^{\infty} \gamma^t r_t$ over a trajectory $\tau = (s_0, a_0, s_1, a_2, ...)$, for the discount factor $\gamma \in (0, 1)$.  To achieve that, the expected return over the discounted rewards $J(\pi) = \mathbb E [R(\tau)]$ is optimized  to obtain the optimal policy $\pi^\star = \mathrm{argmax} \, J(\pi)$.

As a fundamental concept in RL, the value function gives a prediction of the expected, cumulative, discounted, future reward, and provides a measure of how well a given state $s$ or state-action pair $(s, a)$ behaves to generate a higher net return.  The state value $V_{\pi}(s) = \mathbb E[R_t | s_t = s]$ is defined as the expected return for following policy $\pi$ from state $s$. On the other hand,  the action value $Q_{\pi}(s, a) = \mathbb E[R_t | s_t = s, a_t = a]$ is the expected return for selecting action $a$ in state $s$ and then following policy $\pi$. The value functions obey the so-called Bellman equations~\cite{Sutton2018}, which can be solved self-consistently. For example, the action-value function obeys, 
\begin{align}
Q_{\pi}(s, a) = \mathbb E\left[ r(s, a) + \gamma . \max_{a^\prime} Q_\theta (s^\prime, a^\prime) \right].
\end{align}
There are several approaches to optimize the policy, that can be  broadly categorized into three types, namely (a) policy-gradient-based, (b) value-based and (c) actor-critic based methods.  In value-based approaches, such as the Q-learning methods, the policy is optimized to get the net maximum value functions solving the Bellman equations as discussed above. On the other hand, in policy gradient methods the policy parameters are optimized by using gradient descent algorithms:
\begin{align}
\nabla_\theta J (\pi_\theta) = \mathbb E \sum_{t=0}^{T} \left[\nabla_\theta \log \pi_\theta(a_t|s_t) \hat R_t \right],
\end{align}
where $\mathbb E$ represents the expectation value over the trajectory $\tau$.  This basic approach can be improved by using a baseline function, $b(s_t)$,  to reduce the  variance  of  gradient  estimation and  forms the basis of the more advanced state-of-the-art DRL actor-critic algorithms. In the most generalized form, the policy gradient methods work by optimizing the following objective (loss) function:
\begin{equation}
\mathrm {L}^{\rm PG}(\theta) = \hat {\mathbb{E}}_t \left[ \log \pi_\theta(a_\theta|s_t)\hat A_t\right],
\end{equation}
where, $\pi_\theta$ is a stochastic policy, and $\hat{A}_t = Q(s_t, a_t) - V(s_t)$  is an estimator of the advantage function at timestep t, since $R_t$ is an estimate of $Q(a_t, s_t)$.

An actor-critic algorithm learns both a policy and a state-value function, and the value function is used for bootstrapping, i.e., updating a state from subsequent estimates, to reduce variance and accelerate learning~\cite{Sutton2018}, while the critic updates action-value function parameters, and the actor updates policy parameters, in the direction suggested by the critic.

The Soft Actor-Critic (SAC) algorithm is a recently proposed actor-critic algorithm \cite{sac} in the field of reinforcement learning. The main difference of SAC compared to other actor-critic methods is that its policy is optimized in an entropy-regularized manner and is inherently stochastic. The policy is trained to maximize a tradeoff between expected return and entropy, with an increase in entropy leading to more exploration and, in addition, preventing the policy from converging prematurely. The following discussion closely follows the theory presented in~\cite{SpinningUp2021}.

In entropy-regularized RL, the agent gets a bonus reward at each time step proportional to the entropy of the policy, 
\begin{eqnarray}
\pi^* = \arg \max_{\pi} \underset {\tau \sim \pi}{\mathbb E} \left[{ \sum_{t=0}^{\infty} \gamma^t \bigg( R(s_t, a_t, s_{t+1}) + \alpha H\left(\pi(\cdot|s_t)\right) \bigg)}\right],
\end{eqnarray}
where $H(P) = \underset {x \sim P}{\mathbb E} [{-\log P(x)} ]$ denotes the entropy computed from the probability distribution $P$, and $\alpha > 0$ is the trade-off coefficient.  The corresponding value functions in this setting, $V^{\pi}$ and $Q^{\pi}$ are changed to,

\begin{align}
V^{\pi}(s)  = & \underset {\tau \sim \pi} {\mathbb E}\Big[ \sum_{t=0}^{\infty} \gamma^t \big( R(s_t, a_t, s_{t+1}) + \alpha H\left(\pi(\cdot|s_t)\right) \big) | s_0 = s\Big],\\
Q^{\pi}(s,a)  = & \underset {\tau \sim \pi} {\mathbb E}\Big[ \sum_{t=0}^{\infty} \gamma^t R(s_t, a_t, s_{t+1})  + \alpha \sum_{t=1}^{\infty} \gamma^t H\left(\pi(\cdot|s_t)\right)|  s_0 = s, a_0 = a\Big].
\end{align}
With these definitions, $V^{\pi}$ and $Q^{\pi}$ are connected by,
\begin{align}
V^{\pi}(s) = \underset {a \sim \pi}{\mathbb E}\left[{Q^{\pi}(s,a)} + \alpha H\left(\pi(\cdot|s)\right)\right].
\end{align}
The Bellman equation for $Q^{\pi}$ becomes, 
\begin{align}
Q^{\pi}(s,a)  \approx r + \gamma\left(Q^{\pi}(s',\tilde{a}') - \alpha \log \pi(\tilde{a}'|s') \right), \; \tilde{a}' \sim \pi(\cdot|s'),
\end{align}
where, the expectation over the next states, $r$ and the states, $s'$  come from the replay buffer, while the next actions $\tilde{a}'$ are sampled fresh from the policy.  SAC concurrently learns a policy $\pi_{\theta}$ and two Q-functions $Q_{\phi_1}, Q_{\phi_2}$ and set up the loss functions for each Q-function, and takes the minimum Q-value between the two Q approximators,
\begin{align}
L(\phi_i, {\mathcal D}) = \underset{(s,a,r,s',d) \sim {\mathcal D}}{{\mathbb E}}\left[     \Bigg( Q_{\phi_i}(s,a) - y(r,s',d) \Bigg)^2     \right],
\end{align}
where the target is given by
\begin{align}
y(r, s', d) = r + \gamma \left( \min_{j=1,2} Q_{\phi_{\text{targ},j}}(s', \tilde{a}') - \alpha \log \pi_{\theta}(\tilde{a}'|s') \right), \  \tilde{a}' \sim \pi_{\theta}(\cdot|s').
\end{align}

For the studies reported in this article, we design the RL-environments using the framework template of {\em OpenAI-Gym}~\cite{OpenaiGym}. The RL-agents are modelled using SAC policy, with two critic models following the implementation available in the open-source software package {\em Stable-Baselines3}~\cite{stable-baselines3}. For modelling the neural networks and optimization of the network parameters we have used the {\em PyTorch}~\cite{pytorch} ML module.

%%%%%%%%%%%%%%%%%%%%%%%%%%%%%%%%%
\begin{figure}[!hbt]
    \centering
    \includegraphics[trim={0cm 0.0cm 0 0cm},clip, width=0.8\linewidth]{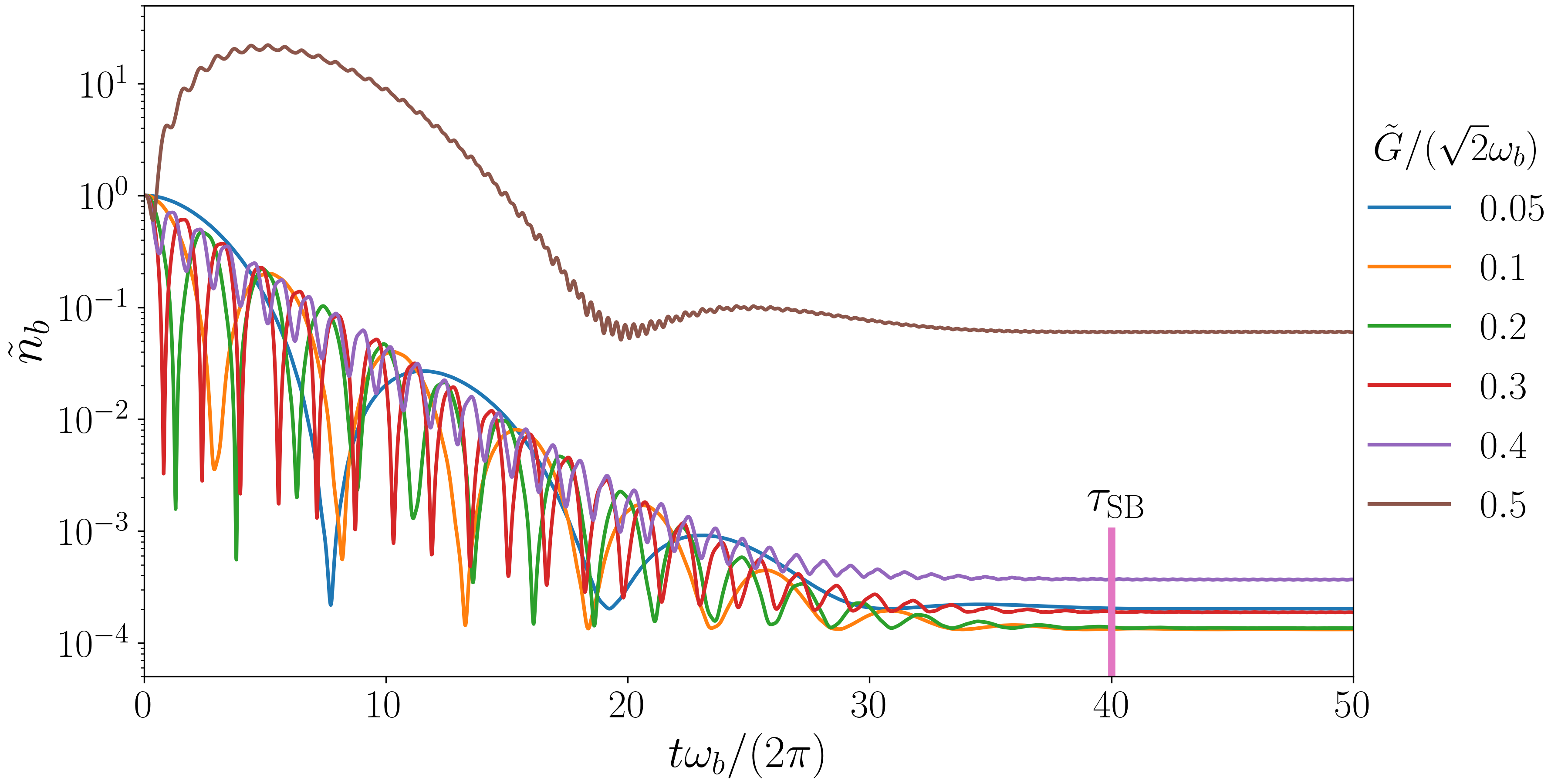}
    \caption{
    This plot demonstrates the mechanical sideband cooling dynamics of the two-mode system with constant magno-mechanical coupling rates given by $\tilde{G}$, for the parameters $\Delta_m/\omega_b=1$, $\kappa_m/\omega_b=0.1$ and $\kappa_b/\omega_b=10^{-5}$. From the figure, one can get an estimate of the steady-state cooling time limit and the optimum cooling quotient, $\tilde{n}_b$. Considering an optimum cooling quotient of, $\tilde{n}_b\approx 10^{-4}$, the best cooling from such sideband cooling process is found to occur at a cooling time limit of the order of $\tau_{\rm {SB}}\omega_b/(2\pi)\approx 40$. 
    }
    \label{fig:sideband-cooling-limit}
\end{figure}
%%%%%%%%%%%%%%%%%%%%%%%%%%%%%%%%%

%%%%%%%%%%%%%%%%%%%%%%%%%%%%%%%%%%%%%%%%%%%%%%%%
\section*{The bipartite magno-mechanical model and sideband cooling time limit}
The effective two-mode magno-mechanical model shown in Fig.~1(b) in the main text consists of a YIG sphere put in a microwave (MW) cavity, where a uniform magnetic field, $B_0$ is applied in the $z$-direction that induces spin wave magnons precessing in the $xy$-plane. The magnetic field component of the cavity mode couples to the magnons through magnetic dipole coupling~\cite{Tabuchi2014Aug,Zhang2014Oct}. The YIG sphere also acts as an excellent mechanical resonator because of the magnetostrictive effect, that gives rise to vibrational acoustic phonon modes~\cite{zhang2016cavity,li2018magnon}. We consider the magnon modes to be driven by a MW field with frequency $\omega_p$ and amplitude $\epsilon_p$~\cite{wang2018bistability,li2018magnon}. The Hamiltonian of the system in a frame rotating with the magnon drive frequency is given by,
\begin{align}
\mathcal{H}_0/\hbar = \delta_a a^\dagger a + \omega_b b^\dagger b + \delta_m m^\dagger m + J(am^\dagger + a^\dagger m) + G m^\dagger m (b + b^\dagger) + i (\epsilon_p m^\dagger - \epsilon^*_p m) ,
\end{align}
where $a(a^\dagger)$, $m(m^\dagger)$ and $b(b^\dagger)$ are the annihilation (creation) operators for the cavity, magnon and phonon modes respectively, $\delta_i = \omega_i - \omega_p$ are the detunings, $J$ is the magnon-photon coupling strength and $G$ is the magno-mechanical coupling rate. The Rabi frequency $\epsilon_p$ is given by $\sqrt{5N}\gamma B_0/4$, where $\gamma/2\pi=28\ \rm{GHz/T}$ is the gyromagnetic ratio, $N=\rho_n V$ is the total number of spins, where $\rho_n=4.22\times 10^{27}\ \rm{m}^{-3}$ is the spin density of YIG and $V$ is the volume of the sphere.

The quantum Langevin equations for the average dynamics of the fluctuation operators are given by
\begin{align}
\nonumber
\dot{a} &= -(i\delta_a + \kappa_a) a - i J m,\\
\dot{m} &= -(i\delta_m + \kappa_m) m - i J a - i \tilde{G} (b + b^\dagger),\\
\nonumber
\dot{b} &= -(i\omega_b + \kappa_b) b - i ({\tilde{G}}^* m + \tilde{G} m^\dagger),
\end{align}
% with the  the operators denote the mean values.
%%%%%%%%%%%%%
where $\tilde{G}$ is the driving-enhanced magno-mechanical coupling. 
When the cavity has a low Q-factor, with a damping rate of the order of $\kappa_a\gg \{J,\ \tilde{G},\ \kappa_m, \kappa_b\}$, the cavity field can be adiabatically eliminated, that gives rise to an effective magno-mechanical Hamiltonian of the form,
\begin{align}
\tilde{\mathcal{H}}/\hbar = \omega_b b^\dagger b + {\Delta}_m m^\dagger m + (\tilde{G} m^\dagger +  \tilde{G}^* m) (b + b^\dagger) ,
\end{align}
where the effective parameters are given by,
$\Delta_m = \delta_m - \frac{|J|^2}{\delta_a^2 + \kappa_a^2} \delta_a$, $\tilde{\kappa}_m = \kappa_m + \frac{|J|^2}{\delta_a^2 + \kappa_a^2} \kappa_a$. 
Hence, when the sphere is put at a node of the cavity magnetic field, or the cavity damping is very high, effectively the system reduces to a two-mode coupled system with a magno-mechanical interaction.

The magnon and mechanical modes have largely different frequencies, with $\omega_m \sim 2\pi \times 10\ {\rm {GHz}}$ and $\omega_b \sim 2\pi \times 10\ {\rm {MHz}}$, therefore the magnon mode behaves as a `cold' sink at zero temperature for the `hot' mechanical mode. At the red sideband, $\Delta_m = \omega_b$, by virtue of the resonant interaction in the weak coupling regime, between the mechanical mode and magnon anti-Stokes sideband, it gives rise to transfer of thermal quanta from the mechanical mode to the magnon mode, which is the working principle of the well-known sideband cooling technique.
In Fig.~\ref{fig:sideband-cooling-limit} we show the sideband cooling dynamics with several values of $\tilde{G}$, in the regime $\tilde{G}<\omega_b$. The time limit for such irreversible cooling is denoted by $\tau_{\rm {SB}}$. 
From the figure, one can get an estimate
of the steady-state cooling time limit and the optimum cooling quotient, $\tilde{n}_b$. 
Considering an optimum cooling quotient of, $\tilde{n}_b\approx 10^{-4}$,
the best steady-state cooling from such sideband cooling mechanism is found to occur at a time limit of the order of $\tau_{\rm {SB}}\omega_b/(2\pi) \approx 40$.
As can be seen from the figure, such sideband cooling is not possible with higher values of the magno-mechanical coupling, $\tilde{G}$ because of the effect of the counter-rotating terms beyond RWA.
In our DRL-scheme, we show that cooling can be obtained even in the strong coupling regime with proper choice of coupling modulations. And more interestingly, it leads to lower time limit for cooling.

\section{The DRL controller for the bipartite magno-mechanical system}
\begin{figure}[!hbt]
    \centering
    \includegraphics[trim={0cm 0.7cm 0 0cm},clip, width=0.8\linewidth]{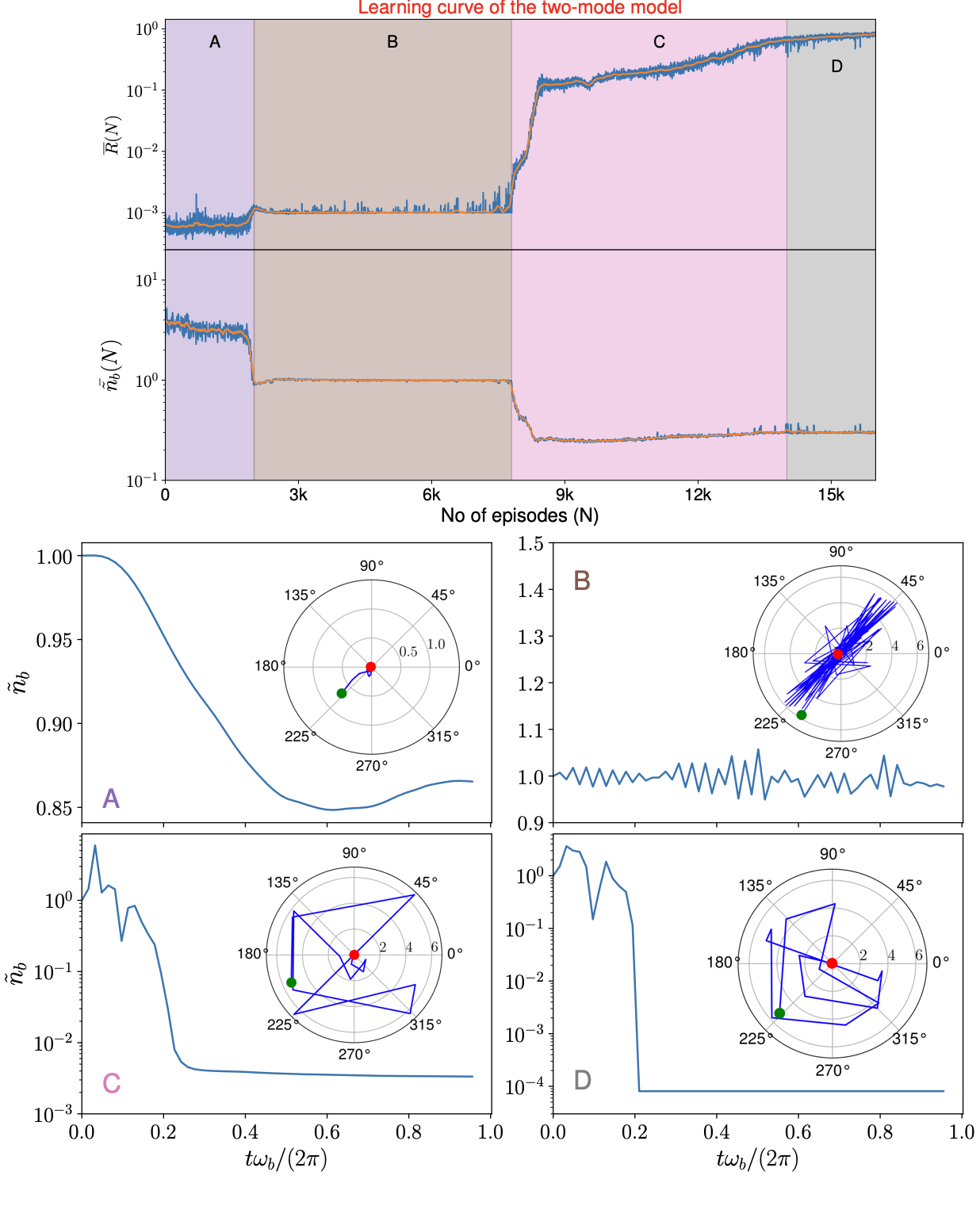}
    \caption{(Top) The learning curve of the RL-agent is shown for the bipartite magno-mechanical model shown in Fig.~1(b) in the main body of the article. In the upper panel of the figure, the mean net reward $\bar{R}(N)$ is shown over each episode $N$ within the training period of the RL agent. The corresponding cooling quotient over each episode $N$ is shown in the lower panel of the figure. To understand the learning process of the RL-agent, we divide the learning curve into four distinct regions, namely A, B, C, and D. The typical performance of the RL agent within these regions is evaluated and plotted respectively in Figs.~A, B, C, and D below the learning curves. In each plot, the variation of cooling quotient, $\tilde{n}_b$ is shown for an episode in which the controls, shown as polar plots in the inset, are determined by the selected RL-agent having already acquired knowledge up to that point of the learning curve. In A and B, the RL agent tries more or less random combinations of the controls and tries to learn the big lessons to achieve a net large return. In A, the RL agent learns to avoid increasing the cooling quotient, while in B it learns to choose controls such that $\tilde{n}_b$ can be held constant at $\tilde{n}_b \sim 1$. The RL agent must explore for a relatively long time to gain control and move beyond it, which does not happen until about $N=8000$. Most of the learning to achieve cooling takes place in region C, where the RL-agent tries to learn the little nitty-gritty of the problem by making its controls more fine-tuned to achieve an overall larger return, which is further improved in the learning period marked by D.}
    \label{fig:two_mode_learning}
\end{figure}

For the two-mode magno-mechanical system, the control parameter is the coupling strength $\tilde G(t)$, which can be complex in nature (see Eq.~1 of the main text of the article). In the DRL model, the real and imaginary parts of $\tilde G$ are considered as the two actions that are learned by the DRL-agent through trial and error. In a given iteration of the RL workflow at time $t$, the RL-agent chooses two actions based on which the system (RL-environment) makes the dynamics to $t = t + \delta t$, where $\delta t = 0.1 \omega_b/(2\pi)$ during which $\tilde G(t)$ remains fixed. The observations that the RL-agent gets from the RL-environment comprise the instantaneous values of the moments obtained by solving the second order coupled differential equations. The actions of the RL-agent are also added to the list of observations, which we found to be advantageous for learning. The SAC controller is made based on the following settings of hyperparameters,
\begin{table}[h]
% \internallinenumbers
\begin{tabular}{ll}
% \internallinenumbers
\small 
Network hidden layer size: & $512 \times 256 \times 256 \times 128$ \\
Activation function: & Rectified Linear Unit (ReLU) \\
Learning rate: & $10^{-4}$ \\
Buffer size: & 1000000 \\ 
Batch size: & 512 \\
Soft update coefficient, $\tau$: & 0.005 \\
Discount factor, $\gamma$: & 0.99 \\
Entropy regularization \\coefficient, $\alpha$: & Adaptive starting from 0.1. \\
\end{tabular}
\end{table}

The goal of the problem is to optimize the neural network parameters such that it learns the non-trivial combinations of the controls, $\tilde G$, so that the phonon occupancy of the YIG sphere is reduced thereby cooling it significantly. For that the agent is given the following reward function,  
\begin{align}
    r(t) ~=~& \frac{n_T}{{\langle b^\dagger b (t)\rangle}},
    \label{eq:reward_function_two_mode}
\end{align} 
where $n_T$ is the thermal phonon number, $b^\dagger b$ is the number operator of the mechanical mode, and $\langle A \rangle$ denotes the expectation value of the operator $A$. The task of the RL-agent would be to maximize the net reward, $R=\sum_i^{{\mathcal N}_t} r_i$ over an episode. Note that this, in theory, represents a highly challenging task as for each control (action), the number of possible combinations in a given episode is $\mathcal{N}_C^{\mathcal{N}_t}$, where $\mathcal{N}_t$ is the number of total time steps allowed within an episode, and ${\mathcal{N}_C}$ is the number of possible discretization of the control strength bounded by $\tilde G_{\rm max}$. Since in our case, the actions are continuous the problem is even harder. 

The RL-agent is trained for the following set of cases: $\tilde{G}_{\rm max}/(\sqrt{2}\omega_b) = \left\{0.1, 0.2, 0.3, 0.5, 1, 2, 3, 4, 5\right\}$. The learning curve for $\tilde{G}_{\rm max}/(\sqrt{2}\omega_b) = 5$ is shown in Fig.~\ref{fig:two_mode_learning}. In these plots, we have shown how the RL-agent gradually learns the non-intuitive control sequences by optimizing the neural network weights to achieve the goal of attaining larger net returns determined by Eq.~\ref{eq:reward_function_two_mode}. We have also shown the performance of the RL-agent at different stages to demonstrate the learning process, as explained in the caption of  Fig.~\ref{fig:two_mode_learning} in more detail.

\begin{figure}[!hbt]
    \centering
    \includegraphics[width=1.0\linewidth]{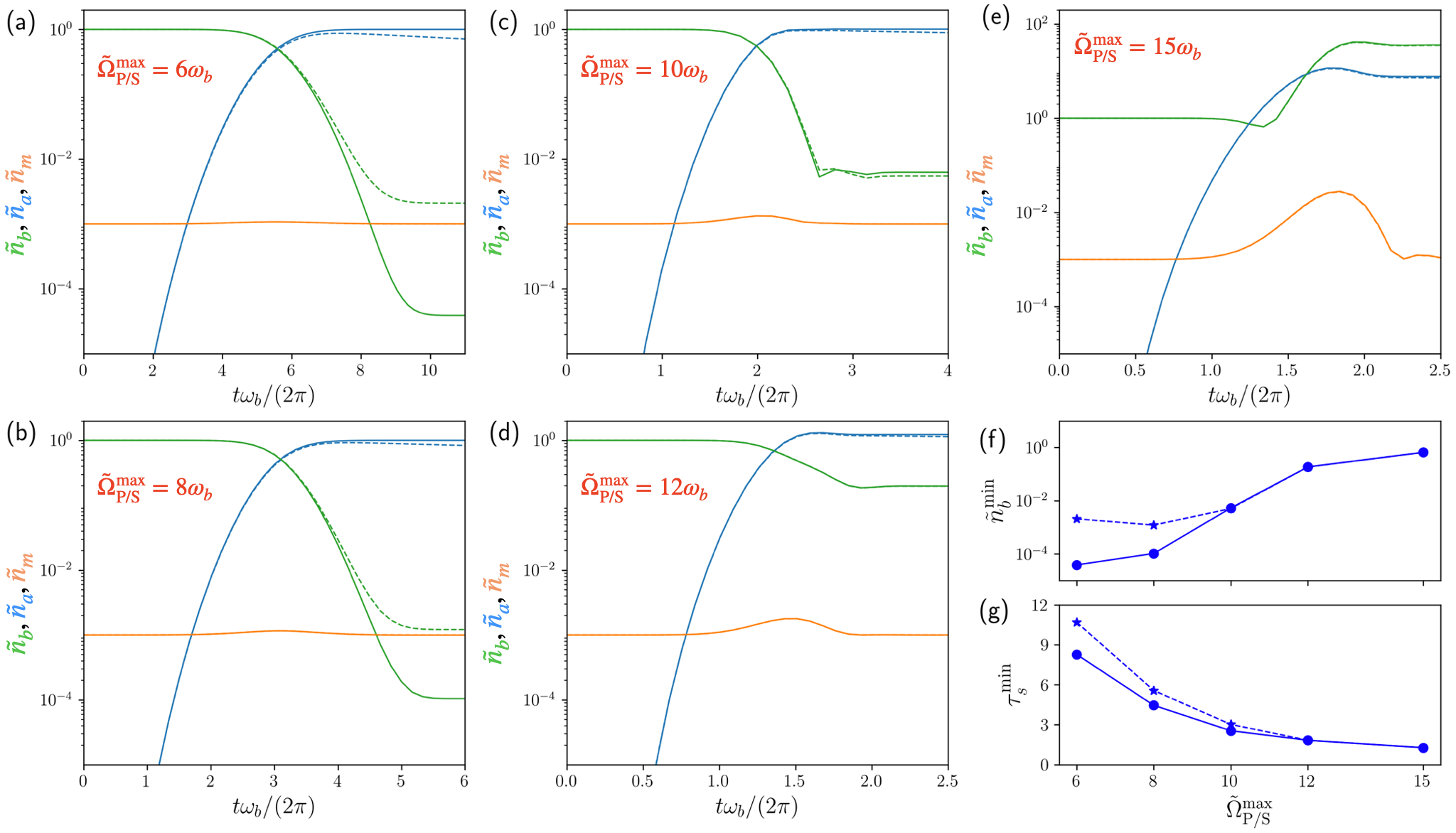}
    \caption{The effectiveness of STIRAP protocol for the tripartite opto-magno-mechanical system for different $\tilde \Omega_{\rm P/S}^{\rm max}$ with numerically optimized counterintuitive Gaussian pulses, typically used in such protocols are compared. The performance with (dotted lines) and without damping (solid lines) are shown. For (a) and (b), such a protocol leads to cooling of the phonon mode ($\tilde n_b$) below 3 orders without damping. For $\tilde \Omega_{\rm P/S}^{\rm max}=10\omega_b$, shown in (c),  and for $\tilde \Omega_{\rm P/S}^{\rm max}=12\omega_b$, in (d), the STIRAP protocol cannot lead to any effective cooling, because of the counter-rotating terms of the non-RWA model. This can be more easily understood for  $\tilde \Omega_{\rm P/S}^{\rm max}=15\omega_b$, shown in (e), where the populations are increased leading to heating of the system. In (f), we compare the minimum achievable phonon occupation, $\tilde n_b^{\rm min}$ for the above choices of $\tilde \Omega_{\rm P/S}^{\rm max}$ with (dotted) and without (solid) damping. In (g), the time required to cool the phonon occupation below $10^{-3}$ or the time required to attain the minimum occupation, $\tau_s^{\rm min}$ is compared for cases with (dotted) and without (solid) damping. }
    \label{fig:three_mode_stirap_limit}
\end{figure}

\section*{Tripartite opto-magno-mechanical model and the cooling time limit}

For the tripartite system model, we consider the cooling of a levitated spherical YIG ferrimagnet with high spin density, trapped in a harmonic potential wherein spin-wave magnon modes are excited by applying an external bias magnetic field.
In a large external homogeneous magnetic field, $B_{0}$, applied in the $\hat{z}$-direction, the YIG sphere is magnetized to its
saturation magnetization, $M_{s}=5.87 \times 10^{5} \mathrm{Am}^{-1}$, whereas the small magnetization fluctuations, $\boldsymbol{m}(\boldsymbol{r}, t)$ around the fully magnetized state is given by, $\boldsymbol{M}(\boldsymbol{r}, t)=M_{S} \hat{z}+\boldsymbol{m}(\boldsymbol{r}, t)$.
For the homogeneously magnetized fundamental magnon mode (Kittel mode), precessing around the $\hat{z}$-axis with an eigen-frequency, $\omega_m=|\gamma| B_{0}$, the magnetization fluctuation is given by $\boldsymbol{m}(\boldsymbol{r}, t)=\mathcal{M}_{K}(\hat{x}+i \hat{y}) m+ h. c .$~\cite{Fletcher1959ferrimagnetic}, which induces a magnetic field,
$
{\boldsymbol{b}}(\boldsymbol{r}, t)=\mu_{0}\frac{\mathcal{M}_{K}}{3} \frac{R^{3}}{r^{3}} e^{i \phi}\left(2 \sin \theta\  \hat{r}-\cos \theta\  \hat{\theta}-i \hat{\phi}\right) m +\textit {h.c.}, 
$
outside the sphere, where, $\mathcal{M}_{K}=\sqrt{\frac{\hbar \gamma M_{s}}{2 V}}$ is the zero-point magnetization, $\mu_0$ is the vacuum permeability, $m$ is the bosonic magnon operator, $R$ and $V$ are the radius and the volume of the sphere.
As schematically shown in the main text, a driven WGM optical microsphere with a magnetostrictive rod (MR) attached to it, is placed near the YIG magnet. 
The magnon mode field modulates the axial length of the MR, 
which modulates the WGM optical mode frequency that is depicted by a coupling of the form $\Omega_S a^\dagger a(m+m^\dagger)$ between the WGM optical mode $(a)$ and the magnon bosonic mode $(m)$, where $\Omega_S=\Delta \omega$ is the optical frequency shift i.e.~the single photon magnon-cavity coupling.

The magnon mode can also be coupled to the COM motion of the YIG sphere, by applying a spatially inhomogeneous external time-dependent magnetic field, $\mathbf{H}_{g}(y, t)$~\cite{Hoang2016electron,Delord2018ramsey}, which satisfies the weak driving, $\left|\mathbf{H}_{g}(y, t)\right| \ll H_{0}$, and small-curl, $\left|\nabla \times \mathbf{H}_{g}(y, t) \| \mathbf{r}\right| \ll\left|\mathbf{H}_{g}(y, t)\right|$ conditions. 
The COM motion of the magnet with frequencies $\omega_{x}, \omega_{y}$, and $\omega_{z}$, in $\hat{x}, \hat{y}$ and $\hat{z}$ directions, is depicted by the Hamiltonian, ${H}_{b} = \sum_{j=x, y, z} \omega_{j} {b}_{j}^{\dagger} {b}_{j}$, where the bosonic ladder operators describe annihilation and creation of a motional quantum along the direction $j$. 
The quantum Hamiltonian describing the interaction between the COM motion, and the spin-wave magnetization due to the gradient magnetic field is given by,
$\mathcal{H}_{mb} (t)=-\frac{\mu_{0}}{2} \int d V\left[M_{S} \hat{z}+{\mathbf{m}}(\mathbf{r})\right] \cdot \mathbf{H}_{g}\left(\mathbf{y}, t\right)$.
%%%
Considering a time-varying gradient magnetic field of the form, $\mathbf{H}_g (\mathbf{y}, t)=\frac{b_g(t)}{\mu_{0}} y \hat{y}$, ($b_g$ in units of $[\mathrm{T} / \mathrm{m}]$), the interaction Hamiltonian for the COM motion in the $\hat{y}$-direction (frequency $\omega_y \equiv \omega_b$ for $b_y \equiv b$) and the magnon mode is given by, 
$
\mathcal{H}_{mb} (t)= \tilde{\Omega}_P(t) \left(\hat{b}+\hat{b}^{\dagger}\right) \left(\hat{m}+\hat{m}^{\dagger}\right), 
$
with $\tilde{\Omega}_P(t) =\frac{b_g (t)}{4} \sqrt{\frac{|\gamma| M_{S}}{\rho \omega_{b}}}$, where $\rho = 5170\ \rm{kg/m^3}$ is the mass density of YIG.
%%%%%%
Taking into account all these interactions, in the rotating frame of the cavity drive and the displacement picture of the average field in each mode~\cite{Wang2012using}, the Hamiltonian is given by
\begin{align}
\nonumber
\tilde{\mathcal{H}}/\hbar = & \Delta_{a}  a^{\dagger } a+\omega_{m} m^{\dagger} m+\omega
_{b} b^{\dagger } b  + \tilde{\Omega}_S(a + a^{\dagger})(m + m^{\dagger }) \\
+& ~ \tilde{\Omega}_P(m + m^{\dagger})(b + b^{\dagger }),  
\end{align}%
where ${\Delta}_{a}$ is the cavity detuning, and $\tilde{\Omega}_S$ is the driven optomagnonic
coupling. 
%
%
%%%%%%%%%%%%%%%%%
\begin{figure}[!hbt]
    \centering
    \includegraphics[width=1.0\linewidth]{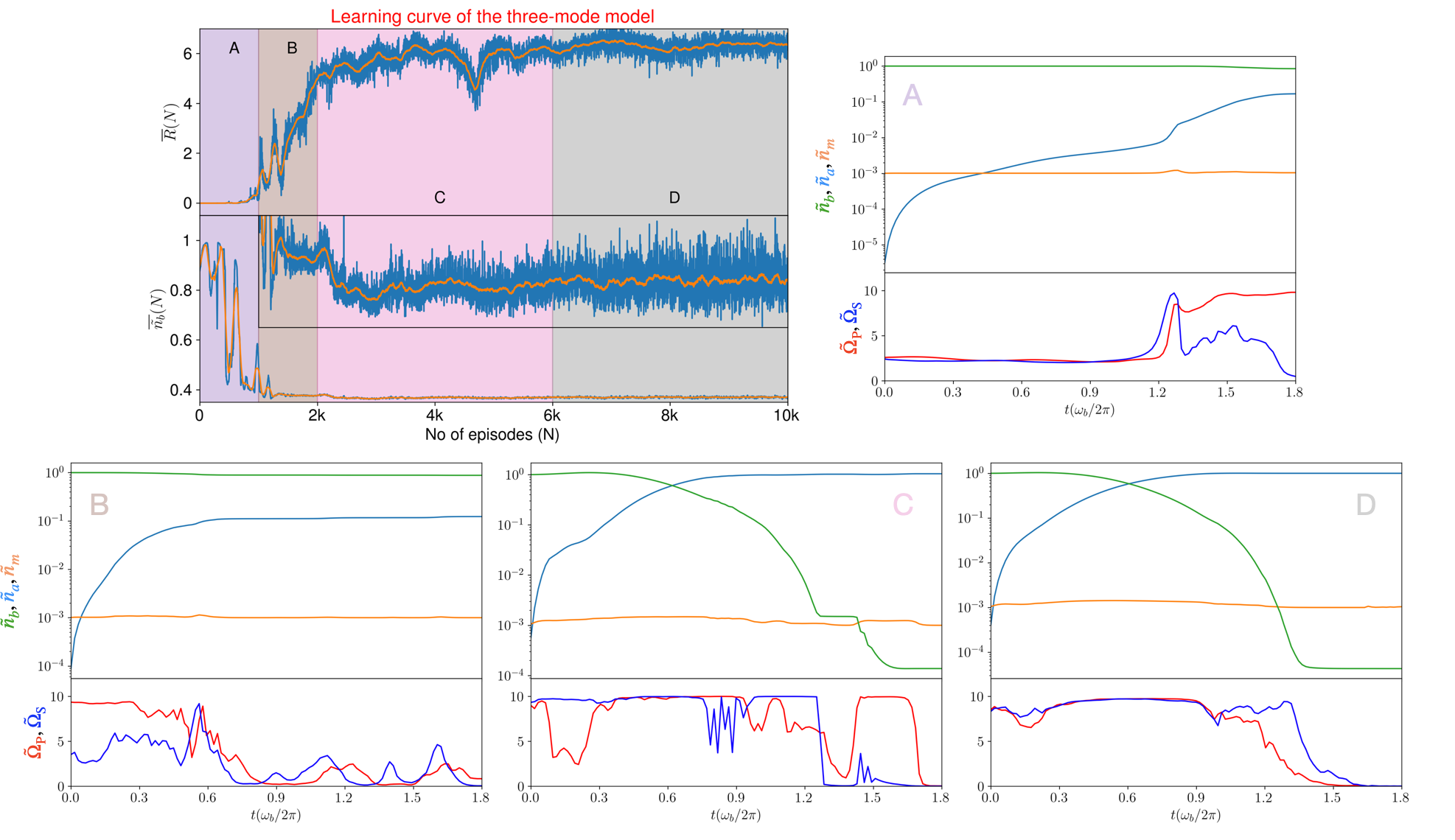}
    \caption{(Top) The learning curve of the RL-agent is shown for the three-mode opto-magno-mechanical model shown in Fig.1(c) in the main text of the paper. In the upper panel of the figure, the mean net reward $\tilde{R}(N)$ (scaled by the thermal number) is shown over each episode $N$ within the training period. The corresponding variation in the cooling quotient of the mean phonon number averaged over each episode, $\bar{\tilde n}_b(N)$ is shown in the lower panel of the figure. The zoomed view of the region beyond $N=1k$ is dubbed in the inset of the plot. To understand the learning process of the RL-agent better, we divide the learning curve into four distinct regions, namely A, B, C, and D. The typical performance of the RL-agent within these regions is evaluated and plotted respectively in Figs. A, B, C, and D.  In each plot, the variation of $\tilde n_b$ (in green), $\tilde n_m$ (in orange)  and $\tilde n_a$ (in blue) are shown for an episode in which the controls, $\tilde \Omega_{\rm P/S}$ are shown in the lower panels (similar to the Fig. 3(b) in the main body of the article), which are determined by the selected RL-agent having already acquired knowledge up to that point of the learning curve. In the region A learning curve, the RL-agent explores in large parameter space more or less randomly to search for controls that would reduce the mean phonon number from within its thermal population limit, $n_b^T$. At the end of stage A, the RL-agent more or less learns that applying constant control pulses cannot lead to a transfer of population out of the phonon mode. In the middle part of B, the RL-agent learns that it is possible to get considerably higher net returns if larger amplitude of control pulses are followed by smaller ones. The RL-agent then learns to apply the smaller controls appropriately at precise times which leads to better cooling through population transfer in the region C and D.}
    \label{fig:learning_three_mode}
\end{figure}
%%%%%%%%%%%%%%%%%%%

In this tripartite system, the optical and COM modes are not directly coupled, however both are coupled to the magnon mode. 
In the large detuning condition, $\omega_m\gg \sqrt{\tilde{\Omega}_P^2 + \tilde{\Omega}_S^2}$, and under RWA, the model can be reduced to a bipartite effective system with the form of Hamiltonian,
\begin{align}
    \tilde{\mathcal{H}}/\hbar = \begin{pmatrix}
-\Delta_{\rm {eff}} &  -\Omega_{\rm{eff}}\\
-\Omega_{\rm{eff}} & \Delta_{\rm {eff}}
\end{pmatrix},
\end{align}
where $\Delta_{\rm{eff}}=(\tilde{\Omega}_P^2 - \tilde{\Omega}_S^2)/(2\omega_m)$ and $\Omega_{\rm {eff}}=\tilde{\Omega}_S \tilde{\Omega}_P/\omega_m$.
This results in a transfer time limit of $\tilde{\tau}_{\rm {lim}}=\pi \omega_m/(2\tilde{\Omega}_S \tilde{\Omega}_P)$.
In the three-mode system given by the full tripartite model under RWA, a widely applicable scheme of population transfer between the uncoupled modes in an irreversible manner is to apply STIRAP-like ideal pulses~\cite{Bergmann1998Jul}. 
However, only when the total operation time of pulse sequences, $T\gg \tilde{\tau}_{\rm {lim}}$, adiabatic effective population transfer can occur between the optical and COM modes in this system.
In Fig.~\ref{fig:three_mode_stirap_limit}, we show the cooling dynamics with ideal STIRAP-like Gaussian pulses. This is obtained with counter-intuitive sequence of the coupling parameters for transfer of quanta from the COM mode to the optical mode. The cooling time is of the order of the limit given from the above expression, $\tilde{\tau}_{\rm {lim}}$, which shows that these pulses are not adiabatic, but rather gives the shortest time for excitation transfer. The performance of the cooling quotient can be seen from the figure. As the time limit is the shortest possible time for cooling, one has to compromise on the effective cooling achieved. For the coupling parameters in the range, $\tilde{\Omega}_i^{\rm {max}}/\omega_b=\{6,8,10\}$, the RWA is valid, whereas for higher values of $\tilde{\Omega}_i^{\rm {max}}$ beyond $12\omega_b$, the counter-rotating terms have significant effect. Therefore, even if the increase of coupling parameters shows sign of the lowering of transfer time, it is not possible to obtain effective cooling in these cases. On the contrary, our proposed DRL-scheme finds finely-tuned pulses to go beyond such limits to obtain effective and much faster cooling.

\begin{figure}[!hbt]
    \centering
    \includegraphics[width=1.0\linewidth]{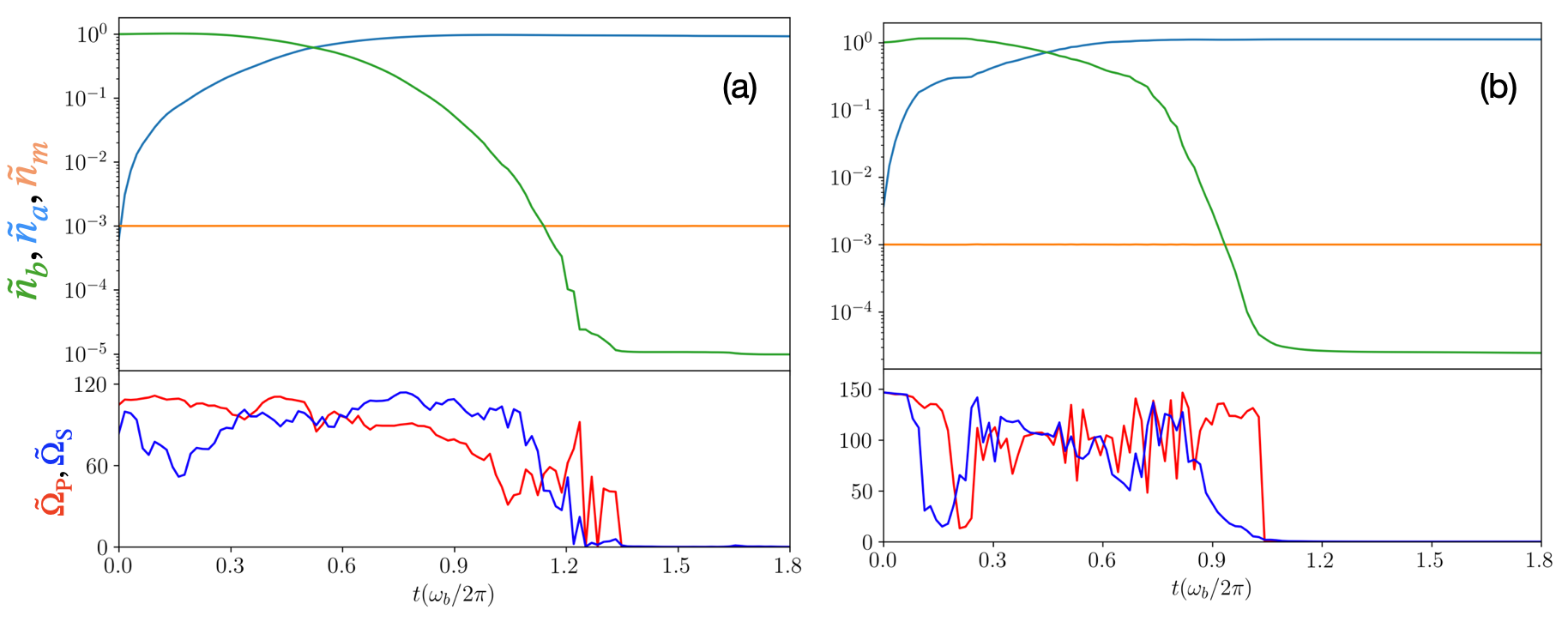}
    \caption{The same as Fig.~3 in the main text of the article, but for the parameters (a) $\tilde{\Omega}_{\rm {S/P}}^{\rm{max}}/\omega_b={120}$, (b) (a) $\tilde{\Omega}_{\rm {S/P}}^{\rm{max}}/\omega_b={150}$. }
    \label{fig:supp_three_mode}
\end{figure}

\section{The DRL controller for the tripartite opto-magno-mechanical system}

The tripartite system is modelled similarly to the two-mode model described earlier, except the fact that an additional Gaussian noise layer is added to the final layer of the network to facilitate more exploration to uncover the non-triviality. Here the controls are, $\tilde \Omega_{\rm P/S}$, which are taken to be real and positive. As earlier, the observables are the instantaneous solutions of the coupled differential equations of the second order moments of the quantum master equation of the tripartite system, along with the controls, $\tilde \Omega_{\rm P/S}$.  In a given episode, the RL-agent makes a total of 150 interactions with the RL-environment in timesteps of $dt=0.1$, each time applying its actions, maintaining a generous exploration-exploitation trade-off. The reward function, for this particular case is,
\begin{align}
    r(t) ~=~& \frac{n_{T}}{{\langle b^\dagger b (t)\rangle}} - \lambda \, {\langle m^\dagger m }(t)\rangle,
    \label{eq:reward_function_three_mode}
\end{align}
where $n_T$ is the thermal phonon number, $b^\dagger b ~(m^\dagger m)$ is the number operator of the phonon (magnon) modes, and $\lambda = 10$ is chosen which guarantees to shuffle population out of the  phonon mode and into the optical cavity mode leading to cooling of the phonon mode, without transferring to the magnon modes.

Given the large computational overhead to solve the coupled second order moment equations for the tripartite opto-magno-mechanical system with state-of-the-art experimental parameter choice of  $\omega_m = 10^5\,\omega_b$, the problem is solved in two steps. In the first step, we train an \emph{auxiliary} system with $\omega_m = 10^3 \omega_b$ and $\tilde \Omega_{\rm P/S}=10\ \omega_b$, for which solution of the set of equations can be obtained without much computational effort. The trained auxiliary model is then used as a supervisor/teacher for the actual system, that we call \emph{primary}, with $\omega_m = 10^5\omega_b$ and $\tilde \Omega_{\rm P/S}=100\ \omega_b$. Training the actual system for a few hundred episodes with periodic evaluation of the RL-agent yields the best trained model. In the literature of RL, this can be considered as a form of \textit{imitation learning}.

The (auxiliary) RL-agent is trained for the following set of cases: $\tilde{\Omega}_{\rm P/S}^{\rm max} = 10, 12, 15$. The learning curve for the first case is shown in Fig.~\ref{fig:learning_three_mode}. In these plots, we have shown how the RL-agent gradually learns the non-intuitive control sequences by optimizing the neural network weights to achieve the goal of attaining larger net returns determined by Eq.~\ref{eq:reward_function_three_mode}. We have also shown the performance of the RL-agent at different stages of the learning process, as explained in the caption of Fig.~\ref{fig:learning_three_mode} in more detail.

%%%%%
% \bibliographystyle{apsrev4-1}
% \bibliography{bibfile}

\begin{thebibliography}{54}%
\makeatletter
\providecommand \@ifxundefined [1]{%
 \@ifx{#1\undefined}
}%
\providecommand \@ifnum [1]{%
 \ifnum #1\expandafter \@firstoftwo
 \else \expandafter \@secondoftwo
 \fi
}%
\providecommand \@ifx [1]{%
 \ifx #1\expandafter \@firstoftwo
 \else \expandafter \@secondoftwo
 \fi
}%
\providecommand \natexlab [1]{#1}%
\providecommand \enquote  [1]{``#1''}%
\providecommand \bibnamefont  [1]{#1}%
\providecommand \bibfnamefont [1]{#1}%
\providecommand \citenamefont [1]{#1}%
\providecommand \href@noop [0]{\@secondoftwo}%
\providecommand \href [0]{\begingroup \@sanitize@url \@href}%
\providecommand \@href[1]{\@@startlink{#1}\@@href}%
\providecommand \@@href[1]{\endgroup#1\@@endlink}%
\providecommand \@sanitize@url [0]{\catcode `\\12\catcode `\$12\catcode
  `\&12\catcode `\#12\catcode `\^12\catcode `\_12\catcode `\%12\relax}%
\providecommand \@@startlink[1]{}%
\providecommand \@@endlink[0]{}%
\providecommand \url  [0]{\begingroup\@sanitize@url \@url }%
\providecommand \@url [1]{\endgroup\@href {#1}{\urlprefix }}%
\providecommand \urlprefix  [0]{URL }%
\providecommand \Eprint [0]{\href }%
\providecommand \doibase [0]{http://dx.doi.org/}%
\providecommand \selectlanguage [0]{\@gobble}%
\providecommand \bibinfo  [0]{\@secondoftwo}%
\providecommand \bibfield  [0]{\@secondoftwo}%
\providecommand \translation [1]{[#1]}%
\providecommand \BibitemOpen [0]{}%
\providecommand \bibitemStop [0]{}%
\providecommand \bibitemNoStop [0]{.\EOS\space}%
\providecommand \EOS [0]{\spacefactor3000\relax}%
\providecommand \BibitemShut  [1]{\csname bibitem#1\endcsname}%
\let\auto@bib@innerbib\@empty
%</preamble>
\bibitem [{\citenamefont {Sch{\ifmmode\ddot{a}\else\"{a}\fi}fermeier}\ \emph
  {et~al.}(2016)\citenamefont {Sch{\ifmmode\ddot{a}\else\"{a}\fi}fermeier},
  \citenamefont {Kerdoncuff}, \citenamefont {Hoff}, \citenamefont {Fu},
  \citenamefont {Huck}, \citenamefont {Bilek}, \citenamefont {Harris},
  \citenamefont {Bowen}, \citenamefont {Gehring},\ and\ \citenamefont
  {Andersen}}]{Schafermeier2016Nov}%
  \BibitemOpen
  \bibfield  {author} {\bibinfo {author} {\bibfnamefont {C.}~\bibnamefont
  {Sch{\ifmmode\ddot{a}\else\"{a}\fi}fermeier}}, \bibinfo {author}
  {\bibfnamefont {H.}~\bibnamefont {Kerdoncuff}}, \bibinfo {author}
  {\bibfnamefont {U.~B.}\ \bibnamefont {Hoff}}, \bibinfo {author}
  {\bibfnamefont {H.}~\bibnamefont {Fu}}, \bibinfo {author} {\bibfnamefont
  {A.}~\bibnamefont {Huck}}, \bibinfo {author} {\bibfnamefont {J.}~\bibnamefont
  {Bilek}}, \bibinfo {author} {\bibfnamefont {G.~I.}\ \bibnamefont {Harris}},
  \bibinfo {author} {\bibfnamefont {W.~P.}\ \bibnamefont {Bowen}}, \bibinfo
  {author} {\bibfnamefont {T.}~\bibnamefont {Gehring}}, \ and\ \bibinfo
  {author} {\bibfnamefont {U.~L.}\ \bibnamefont {Andersen}},\ }\href {\doibase
  10.1038/ncomms13628} {\bibfield  {journal} {\bibinfo  {journal} {Nat.
  Commun.}\ }\textbf {\bibinfo {volume} {7}},\ \bibinfo {pages} {1} (\bibinfo
  {year} {2016})}\BibitemShut {NoStop}%
\bibitem [{\citenamefont {Guo}\ \emph {et~al.}(2019)\citenamefont {Guo},
  \citenamefont {Norte},\ and\ \citenamefont
  {Gr{\ifmmode\ddot{o}\else\"{o}\fi}blacher}}]{Guo2019Nov}%
  \BibitemOpen
  \bibfield  {author} {\bibinfo {author} {\bibfnamefont {J.}~\bibnamefont
  {Guo}}, \bibinfo {author} {\bibfnamefont {R.}~\bibnamefont {Norte}}, \ and\
  \bibinfo {author} {\bibfnamefont {S.}~\bibnamefont
  {Gr{\ifmmode\ddot{o}\else\"{o}\fi}blacher}},\ }\href {\doibase
  10.1103/PhysRevLett.123.223602} {\bibfield  {journal} {\bibinfo  {journal}
  {Phys. Rev. Lett.}\ }\textbf {\bibinfo {volume} {123}},\ \bibinfo {pages}
  {223602} (\bibinfo {year} {2019})}\BibitemShut {NoStop}%
\bibitem [{\citenamefont {Park}\ and\ \citenamefont
  {Wang}(2009)}]{Park2009Jul}%
  \BibitemOpen
  \bibfield  {author} {\bibinfo {author} {\bibfnamefont {Y.-S.}\ \bibnamefont
  {Park}}\ and\ \bibinfo {author} {\bibfnamefont {H.}~\bibnamefont {Wang}},\
  }\href {\doibase 10.1038/nphys1303} {\bibfield  {journal} {\bibinfo
  {journal} {Nat. Phys.}\ }\textbf {\bibinfo {volume} {5}},\ \bibinfo {pages}
  {489} (\bibinfo {year} {2009})}\BibitemShut {NoStop}%
\bibitem [{\citenamefont {Clark}\ \emph {et~al.}(2017)\citenamefont {Clark},
  \citenamefont {Lecocq}, \citenamefont {Simmonds}, \citenamefont {Aumentado},\
  and\ \citenamefont {Teufel}}]{Clark2017Jan}%
  \BibitemOpen
  \bibfield  {author} {\bibinfo {author} {\bibfnamefont {J.~B.}\ \bibnamefont
  {Clark}}, \bibinfo {author} {\bibfnamefont {F.}~\bibnamefont {Lecocq}},
  \bibinfo {author} {\bibfnamefont {R.~W.}\ \bibnamefont {Simmonds}}, \bibinfo
  {author} {\bibfnamefont {J.}~\bibnamefont {Aumentado}}, \ and\ \bibinfo
  {author} {\bibfnamefont {J.~D.}\ \bibnamefont {Teufel}},\ }\href {\doibase
  10.1038/nature20604} {\bibfield  {journal} {\bibinfo  {journal} {Nature}\
  }\textbf {\bibinfo {volume} {541}},\ \bibinfo {pages} {191} (\bibinfo {year}
  {2017})}\BibitemShut {NoStop}%
\bibitem [{\citenamefont {Frimmer}\ \emph {et~al.}(2016)\citenamefont
  {Frimmer}, \citenamefont {Gieseler},\ and\ \citenamefont
  {Novotny}}]{Frimmer2016Oct}%
  \BibitemOpen
  \bibfield  {author} {\bibinfo {author} {\bibfnamefont {M.}~\bibnamefont
  {Frimmer}}, \bibinfo {author} {\bibfnamefont {J.}~\bibnamefont {Gieseler}}, \
  and\ \bibinfo {author} {\bibfnamefont {L.}~\bibnamefont {Novotny}},\ }\href
  {\doibase 10.1103/PhysRevLett.117.163601} {\bibfield  {journal} {\bibinfo
  {journal} {Phys. Rev. Lett.}\ }\textbf {\bibinfo {volume} {117}},\ \bibinfo
  {pages} {163601} (\bibinfo {year} {2016})}\BibitemShut {NoStop}%
\bibitem [{\citenamefont {Abdi}\ \emph {et~al.}(2016)\citenamefont {Abdi},
  \citenamefont {Degenfeld-Schonburg}, \citenamefont {Sameti}, \citenamefont
  {Navarrete-Benlloch},\ and\ \citenamefont {Hartmann}}]{abdi2016dissipative}%
  \BibitemOpen
  \bibfield  {author} {\bibinfo {author} {\bibfnamefont {M.}~\bibnamefont
  {Abdi}}, \bibinfo {author} {\bibfnamefont {P.}~\bibnamefont
  {Degenfeld-Schonburg}}, \bibinfo {author} {\bibfnamefont {M.}~\bibnamefont
  {Sameti}}, \bibinfo {author} {\bibfnamefont {C.}~\bibnamefont
  {Navarrete-Benlloch}}, \ and\ \bibinfo {author} {\bibfnamefont {M.~J.}\
  \bibnamefont {Hartmann}},\ }\href {\doibase
  https://doi.org/10.1103/PhysRevLett.116.233604} {\bibfield  {journal}
  {\bibinfo  {journal} {Phys. Rev. Lett.}\ }\textbf {\bibinfo {volume} {116}},\
  \bibinfo {pages} {233604} (\bibinfo {year} {2016})}\BibitemShut {NoStop}%
\bibitem [{\citenamefont {Bose}\ \emph {et~al.}(1999)\citenamefont {Bose},
  \citenamefont {Jacobs},\ and\ \citenamefont {Knight}}]{Bose1999May}%
  \BibitemOpen
  \bibfield  {author} {\bibinfo {author} {\bibfnamefont {S.}~\bibnamefont
  {Bose}}, \bibinfo {author} {\bibfnamefont {K.}~\bibnamefont {Jacobs}}, \ and\
  \bibinfo {author} {\bibfnamefont {P.~L.}\ \bibnamefont {Knight}},\ }\href
  {\doibase 10.1103/PhysRevA.59.3204} {\bibfield  {journal} {\bibinfo
  {journal} {Phys. Rev. A}\ }\textbf {\bibinfo {volume} {59}},\ \bibinfo
  {pages} {3204} (\bibinfo {year} {1999})}\BibitemShut {NoStop}%
\bibitem [{\citenamefont {Marshall}\ \emph {et~al.}(2003)\citenamefont
  {Marshall}, \citenamefont {Simon}, \citenamefont {Penrose},\ and\
  \citenamefont {Bouwmeester}}]{Marshall2003Sep}%
  \BibitemOpen
  \bibfield  {author} {\bibinfo {author} {\bibfnamefont {W.}~\bibnamefont
  {Marshall}}, \bibinfo {author} {\bibfnamefont {C.}~\bibnamefont {Simon}},
  \bibinfo {author} {\bibfnamefont {R.}~\bibnamefont {Penrose}}, \ and\
  \bibinfo {author} {\bibfnamefont {D.}~\bibnamefont {Bouwmeester}},\ }\href
  {\doibase 10.1103/PhysRevLett.91.130401} {\bibfield  {journal} {\bibinfo
  {journal} {Phys. Rev. Lett.}\ }\textbf {\bibinfo {volume} {91}},\ \bibinfo
  {pages} {130401} (\bibinfo {year} {2003})}\BibitemShut {NoStop}%
\bibitem [{\citenamefont {Schliesser}\ \emph {et~al.}(2009)\citenamefont
  {Schliesser}, \citenamefont {Arcizet}, \citenamefont
  {Rivi{\ifmmode\grave{e}\else\`{e}\fi}re}, \citenamefont {Anetsberger},\ and\
  \citenamefont {Kippenberg}}]{Schliesser2009Jul}%
  \BibitemOpen
  \bibfield  {author} {\bibinfo {author} {\bibfnamefont {A.}~\bibnamefont
  {Schliesser}}, \bibinfo {author} {\bibfnamefont {O.}~\bibnamefont {Arcizet}},
  \bibinfo {author} {\bibfnamefont {R.}~\bibnamefont
  {Rivi{\ifmmode\grave{e}\else\`{e}\fi}re}}, \bibinfo {author} {\bibfnamefont
  {G.}~\bibnamefont {Anetsberger}}, \ and\ \bibinfo {author} {\bibfnamefont
  {T.~J.}\ \bibnamefont {Kippenberg}},\ }\href {\doibase 10.1038/nphys1304}
  {\bibfield  {journal} {\bibinfo  {journal} {Nat. Phys.}\ }\textbf {\bibinfo
  {volume} {5}},\ \bibinfo {pages} {509} (\bibinfo {year} {2009})}\BibitemShut
  {NoStop}%
\bibitem [{\citenamefont {Manley}\ \emph {et~al.}(2021)\citenamefont {Manley},
  \citenamefont {Chowdhury}, \citenamefont {Grin}, \citenamefont {Singh},\ and\
  \citenamefont {Wilson}}]{Manley2021Feb}%
  \BibitemOpen
  \bibfield  {author} {\bibinfo {author} {\bibfnamefont {J.}~\bibnamefont
  {Manley}}, \bibinfo {author} {\bibfnamefont {M.~D.}\ \bibnamefont
  {Chowdhury}}, \bibinfo {author} {\bibfnamefont {D.}~\bibnamefont {Grin}},
  \bibinfo {author} {\bibfnamefont {S.}~\bibnamefont {Singh}}, \ and\ \bibinfo
  {author} {\bibfnamefont {D.~J.}\ \bibnamefont {Wilson}},\ }\href {\doibase
  10.1103/PhysRevLett.126.061301} {\bibfield  {journal} {\bibinfo  {journal}
  {Phys. Rev. Lett.}\ }\textbf {\bibinfo {volume} {126}},\ \bibinfo {pages}
  {061301} (\bibinfo {year} {2021})}\BibitemShut {NoStop}%
\bibitem [{\citenamefont {Bourassa}\ \emph {et~al.}(2021)\citenamefont
  {Bourassa}, \citenamefont {Quesada}, \citenamefont {Tzitrin}, \citenamefont
  {Sz{\'a}va}, \citenamefont {Isacsson}, \citenamefont {Izaac}, \citenamefont
  {Sabapathy}, \citenamefont {Dauphinais},\ and\ \citenamefont
  {Dhand}}]{bourassa2021fast}%
  \BibitemOpen
  \bibfield  {author} {\bibinfo {author} {\bibfnamefont {J.~E.}\ \bibnamefont
  {Bourassa}}, \bibinfo {author} {\bibfnamefont {N.}~\bibnamefont {Quesada}},
  \bibinfo {author} {\bibfnamefont {I.}~\bibnamefont {Tzitrin}}, \bibinfo
  {author} {\bibfnamefont {A.}~\bibnamefont {Sz{\'a}va}}, \bibinfo {author}
  {\bibfnamefont {T.}~\bibnamefont {Isacsson}}, \bibinfo {author}
  {\bibfnamefont {J.}~\bibnamefont {Izaac}}, \bibinfo {author} {\bibfnamefont
  {K.~K.}\ \bibnamefont {Sabapathy}}, \bibinfo {author} {\bibfnamefont
  {G.}~\bibnamefont {Dauphinais}}, \ and\ \bibinfo {author} {\bibfnamefont
  {I.}~\bibnamefont {Dhand}},\ }\href {\doibase
  https://doi.org/10.1103/PRXQuantum.2.040315} {\bibfield  {journal} {\bibinfo
  {journal} {PRX Quantum}\ }\textbf {\bibinfo {volume} {2}},\ \bibinfo {pages}
  {040315} (\bibinfo {year} {2021})}\BibitemShut {NoStop}%
\bibitem [{\citenamefont {Zhang}\ \emph {et~al.}(2016)\citenamefont {Zhang},
  \citenamefont {Zou}, \citenamefont {Jiang},\ and\ \citenamefont
  {Tang}}]{zhang2016cavity}%
  \BibitemOpen
  \bibfield  {author} {\bibinfo {author} {\bibfnamefont {X.}~\bibnamefont
  {Zhang}}, \bibinfo {author} {\bibfnamefont {C.-L.}\ \bibnamefont {Zou}},
  \bibinfo {author} {\bibfnamefont {L.}~\bibnamefont {Jiang}}, \ and\ \bibinfo
  {author} {\bibfnamefont {H.~X.}\ \bibnamefont {Tang}},\ }\href {\doibase
  https://doi.org/10.1126/sciadv.1501286} {\bibfield  {journal} {\bibinfo
  {journal} {Sci. Adv.}\ }\textbf {\bibinfo {volume} {2}},\ \bibinfo {pages}
  {e1501286} (\bibinfo {year} {2016})}\BibitemShut {NoStop}%
\bibitem [{\citenamefont {Lachance-Quirion}\ \emph {et~al.}(2019)\citenamefont
  {Lachance-Quirion}, \citenamefont {Tabuchi}, \citenamefont {Gloppe},
  \citenamefont {Usami},\ and\ \citenamefont {Nakamura}}]{lachance2019hybrid}%
  \BibitemOpen
  \bibfield  {author} {\bibinfo {author} {\bibfnamefont {D.}~\bibnamefont
  {Lachance-Quirion}}, \bibinfo {author} {\bibfnamefont {Y.}~\bibnamefont
  {Tabuchi}}, \bibinfo {author} {\bibfnamefont {A.}~\bibnamefont {Gloppe}},
  \bibinfo {author} {\bibfnamefont {K.}~\bibnamefont {Usami}}, \ and\ \bibinfo
  {author} {\bibfnamefont {Y.}~\bibnamefont {Nakamura}},\ }\href {\doibase
  https://doi.org/10.7567/1882-0786/ab248d} {\bibfield  {journal} {\bibinfo
  {journal} {Appl. Phys. Express}\ }\textbf {\bibinfo {volume} {12}},\ \bibinfo
  {pages} {070101} (\bibinfo {year} {2019})}\BibitemShut {NoStop}%
\bibitem [{\citenamefont {Wang}\ and\ \citenamefont
  {Hu}(2020)}]{wang2020dissipative}%
  \BibitemOpen
  \bibfield  {author} {\bibinfo {author} {\bibfnamefont {Y.-P.}\ \bibnamefont
  {Wang}}\ and\ \bibinfo {author} {\bibfnamefont {C.-M.}\ \bibnamefont {Hu}},\
  }\href {\doibase https://doi.org/10.1063/1.5144202} {\bibfield  {journal}
  {\bibinfo  {journal} {J. Appl. Phys.}\ }\textbf {\bibinfo {volume} {127}},\
  \bibinfo {pages} {130901} (\bibinfo {year} {2020})}\BibitemShut {NoStop}%
\bibitem [{\citenamefont {Li}\ \emph {et~al.}(2018)\citenamefont {Li},
  \citenamefont {Zhu},\ and\ \citenamefont {Agarwal}}]{li2018magnon}%
  \BibitemOpen
  \bibfield  {author} {\bibinfo {author} {\bibfnamefont {J.}~\bibnamefont
  {Li}}, \bibinfo {author} {\bibfnamefont {S.-Y.}\ \bibnamefont {Zhu}}, \ and\
  \bibinfo {author} {\bibfnamefont {G.}~\bibnamefont {Agarwal}},\ }\href
  {\doibase https://doi.org/10.1103/PhysRevLett.121.203601} {\bibfield
  {journal} {\bibinfo  {journal} {Phys. Rev. Lett.}\ }\textbf {\bibinfo
  {volume} {121}},\ \bibinfo {pages} {203601} (\bibinfo {year}
  {2018})}\BibitemShut {NoStop}%
\bibitem [{\citenamefont {Zhang}\ \emph {et~al.}(2014)\citenamefont {Zhang},
  \citenamefont {Zou}, \citenamefont {Jiang},\ and\ \citenamefont
  {Tang}}]{Zhang2014Oct}%
  \BibitemOpen
  \bibfield  {author} {\bibinfo {author} {\bibfnamefont {X.}~\bibnamefont
  {Zhang}}, \bibinfo {author} {\bibfnamefont {C.-L.}\ \bibnamefont {Zou}},
  \bibinfo {author} {\bibfnamefont {L.}~\bibnamefont {Jiang}}, \ and\ \bibinfo
  {author} {\bibfnamefont {H.~X.}\ \bibnamefont {Tang}},\ }\href {\doibase
  10.1103/PhysRevLett.113.156401} {\bibfield  {journal} {\bibinfo  {journal}
  {Phys. Rev. Lett.}\ }\textbf {\bibinfo {volume} {113}},\ \bibinfo {pages}
  {156401} (\bibinfo {year} {2014})}\BibitemShut {NoStop}%
\bibitem [{\citenamefont {Wang}\ \emph {et~al.}(2018)\citenamefont {Wang},
  \citenamefont {Zhang}, \citenamefont {Zhang}, \citenamefont {Li},
  \citenamefont {Hu},\ and\ \citenamefont {You}}]{wang2018bistability}%
  \BibitemOpen
  \bibfield  {author} {\bibinfo {author} {\bibfnamefont {Y.-P.}\ \bibnamefont
  {Wang}}, \bibinfo {author} {\bibfnamefont {G.-Q.}\ \bibnamefont {Zhang}},
  \bibinfo {author} {\bibfnamefont {D.}~\bibnamefont {Zhang}}, \bibinfo
  {author} {\bibfnamefont {T.-F.}\ \bibnamefont {Li}}, \bibinfo {author}
  {\bibfnamefont {C.-M.}\ \bibnamefont {Hu}}, \ and\ \bibinfo {author}
  {\bibfnamefont {J.}~\bibnamefont {You}},\ }\href {\doibase
  https://doi.org/10.1103/PhysRevLett.120.057202} {\bibfield  {journal}
  {\bibinfo  {journal} {Phys. Rev. Lett.}\ }\textbf {\bibinfo {volume} {120}},\
  \bibinfo {pages} {057202} (\bibinfo {year} {2018})}\BibitemShut {NoStop}%
\bibitem [{\citenamefont {Joshi}\ \emph {et~al.}(2021)\citenamefont {Joshi},
  \citenamefont {Noh},\ and\ \citenamefont {Gao}}]{joshi2021quantum}%
  \BibitemOpen
  \bibfield  {author} {\bibinfo {author} {\bibfnamefont {A.}~\bibnamefont
  {Joshi}}, \bibinfo {author} {\bibfnamefont {K.}~\bibnamefont {Noh}}, \ and\
  \bibinfo {author} {\bibfnamefont {Y.~Y.}\ \bibnamefont {Gao}},\ }\href
  {\doibase https://doi.org/10.1088/2058-9565/abe989} {\bibfield  {journal}
  {\bibinfo  {journal} {Quant. Sci. Tech.}\ }\textbf {\bibinfo {volume} {6}},\
  \bibinfo {pages} {033001} (\bibinfo {year} {2021})}\BibitemShut {NoStop}%
\bibitem [{\citenamefont {Goodfellow}\ \emph {et~al.}(2016)\citenamefont
  {Goodfellow}, \citenamefont {Bengio}, \citenamefont {Courville},\ and\
  \citenamefont {Bengio}}]{Goodfellow2016}%
  \BibitemOpen
  \bibfield  {author} {\bibinfo {author} {\bibfnamefont {I.}~\bibnamefont
  {Goodfellow}}, \bibinfo {author} {\bibfnamefont {Y.}~\bibnamefont {Bengio}},
  \bibinfo {author} {\bibfnamefont {A.}~\bibnamefont {Courville}}, \ and\
  \bibinfo {author} {\bibfnamefont {Y.}~\bibnamefont {Bengio}},\ }\href@noop {}
  {\emph {\bibinfo {title} {Deep learning}}},\ Vol.~\bibinfo {volume} {1}\
  (\bibinfo  {publisher} {MIT press Cambridge},\ \bibinfo {year}
  {2016})\BibitemShut {NoStop}%
\bibitem [{\citenamefont {Sutton}\ and\ \citenamefont
  {Barto}(2018)}]{Sutton2018}%
  \BibitemOpen
  \bibfield  {author} {\bibinfo {author} {\bibfnamefont {R.~S.}\ \bibnamefont
  {Sutton}}\ and\ \bibinfo {author} {\bibfnamefont {A.~G.}\ \bibnamefont
  {Barto}},\ }\href@noop {} {\emph {\bibinfo {title} {{Reinforcement Learning:
  An Introduction}}}},\ \bibinfo {edition} {2nd}\ ed.\ (\bibinfo  {publisher}
  {The MIT Press},\ \bibinfo {year} {2018})\BibitemShut {NoStop}%
\bibitem [{\citenamefont {Silver}\ \emph {et~al.}(2016)\citenamefont {Silver},
  \citenamefont {Huang}, \citenamefont {Maddison}, \citenamefont {Guez},
  \citenamefont {Sifre}, \citenamefont {van~den Driessche}, \citenamefont
  {Schrittwieser}, \citenamefont {Antonoglou}, \citenamefont {Panneershelvam},
  \citenamefont {Lanctot}, \citenamefont {Dieleman}, \citenamefont {Grewe},
  \citenamefont {Nham}, \citenamefont {Kalchbrenner}, \citenamefont
  {Sutskever}, \citenamefont {Lillicrap}, \citenamefont {Leach}, \citenamefont
  {Kavukcuoglu}, \citenamefont {Graepel},\ and\ \citenamefont
  {Hassabis}}]{silver_mastering_2016}%
  \BibitemOpen
  \bibfield  {author} {\bibinfo {author} {\bibfnamefont {D.}~\bibnamefont
  {Silver}}, \bibinfo {author} {\bibfnamefont {A.}~\bibnamefont {Huang}},
  \bibinfo {author} {\bibfnamefont {C.~J.}\ \bibnamefont {Maddison}}, \bibinfo
  {author} {\bibfnamefont {A.}~\bibnamefont {Guez}}, \bibinfo {author}
  {\bibfnamefont {L.}~\bibnamefont {Sifre}}, \bibinfo {author} {\bibfnamefont
  {G.}~\bibnamefont {van~den Driessche}}, \bibinfo {author} {\bibfnamefont
  {J.}~\bibnamefont {Schrittwieser}}, \bibinfo {author} {\bibfnamefont
  {I.}~\bibnamefont {Antonoglou}}, \bibinfo {author} {\bibfnamefont
  {V.}~\bibnamefont {Panneershelvam}}, \bibinfo {author} {\bibfnamefont
  {M.}~\bibnamefont {Lanctot}}, \bibinfo {author} {\bibfnamefont
  {S.}~\bibnamefont {Dieleman}}, \bibinfo {author} {\bibfnamefont
  {D.}~\bibnamefont {Grewe}}, \bibinfo {author} {\bibfnamefont
  {J.}~\bibnamefont {Nham}}, \bibinfo {author} {\bibfnamefont {N.}~\bibnamefont
  {Kalchbrenner}}, \bibinfo {author} {\bibfnamefont {I.}~\bibnamefont
  {Sutskever}}, \bibinfo {author} {\bibfnamefont {T.}~\bibnamefont
  {Lillicrap}}, \bibinfo {author} {\bibfnamefont {M.}~\bibnamefont {Leach}},
  \bibinfo {author} {\bibfnamefont {K.}~\bibnamefont {Kavukcuoglu}}, \bibinfo
  {author} {\bibfnamefont {T.}~\bibnamefont {Graepel}}, \ and\ \bibinfo
  {author} {\bibfnamefont {D.}~\bibnamefont {Hassabis}},\ }\href {\doibase
  10.1038/nature16961} {\bibfield  {journal} {\bibinfo  {journal} {Nature}\
  }\textbf {\bibinfo {volume} {529}} (\bibinfo {year} {2016}),\
  10.1038/nature16961}\BibitemShut {NoStop}%
\bibitem [{\citenamefont {Silver}\ \emph {et~al.}(2017)\citenamefont {Silver},
  \citenamefont {Schrittwieser}, \citenamefont {Simonyan}, \citenamefont
  {Antonoglou}, \citenamefont {Huang}, \citenamefont {Guez}, \citenamefont
  {Hubert}, \citenamefont {Baker}, \citenamefont {Lai}, \citenamefont {Bolton},
  \citenamefont {Chen}, \citenamefont {Lillicrap}, \citenamefont {Hui},
  \citenamefont {Sifre}, \citenamefont {van~den Driessche}, \citenamefont
  {Graepel},\ and\ \citenamefont {Hassabis}}]{silver_mastering_2017}%
  \BibitemOpen
  \bibfield  {author} {\bibinfo {author} {\bibfnamefont {D.}~\bibnamefont
  {Silver}}, \bibinfo {author} {\bibfnamefont {J.}~\bibnamefont
  {Schrittwieser}}, \bibinfo {author} {\bibfnamefont {K.}~\bibnamefont
  {Simonyan}}, \bibinfo {author} {\bibfnamefont {I.}~\bibnamefont
  {Antonoglou}}, \bibinfo {author} {\bibfnamefont {A.}~\bibnamefont {Huang}},
  \bibinfo {author} {\bibfnamefont {A.}~\bibnamefont {Guez}}, \bibinfo {author}
  {\bibfnamefont {T.}~\bibnamefont {Hubert}}, \bibinfo {author} {\bibfnamefont
  {L.}~\bibnamefont {Baker}}, \bibinfo {author} {\bibfnamefont
  {M.}~\bibnamefont {Lai}}, \bibinfo {author} {\bibfnamefont {A.}~\bibnamefont
  {Bolton}}, \bibinfo {author} {\bibfnamefont {Y.}~\bibnamefont {Chen}},
  \bibinfo {author} {\bibfnamefont {T.}~\bibnamefont {Lillicrap}}, \bibinfo
  {author} {\bibfnamefont {F.}~\bibnamefont {Hui}}, \bibinfo {author}
  {\bibfnamefont {L.}~\bibnamefont {Sifre}}, \bibinfo {author} {\bibfnamefont
  {G.}~\bibnamefont {van~den Driessche}}, \bibinfo {author} {\bibfnamefont
  {T.}~\bibnamefont {Graepel}}, \ and\ \bibinfo {author} {\bibfnamefont
  {D.}~\bibnamefont {Hassabis}},\ }\href {\doibase 10.1038/nature24270}
  {\bibfield  {journal} {\bibinfo  {journal} {Nature}\ }\textbf {\bibinfo
  {volume} {550}},\ \bibinfo {pages} {354} (\bibinfo {year}
  {2017})}\BibitemShut {NoStop}%
\bibitem [{\citenamefont {Carleo}\ \emph {et~al.}(2019)\citenamefont {Carleo},
  \citenamefont {Cirac}, \citenamefont {Cranmer}, \citenamefont {Daudet},
  \citenamefont {Schuld}, \citenamefont {Tishby}, \citenamefont
  {Vogt-Maranto},\ and\ \citenamefont
  {Zdeborov{\ifmmode\acute{a}\else\'{a}\fi}}}]{Cirac2019}%
  \BibitemOpen
  \bibfield  {author} {\bibinfo {author} {\bibfnamefont {G.}~\bibnamefont
  {Carleo}}, \bibinfo {author} {\bibfnamefont {I.}~\bibnamefont {Cirac}},
  \bibinfo {author} {\bibfnamefont {K.}~\bibnamefont {Cranmer}}, \bibinfo
  {author} {\bibfnamefont {L.}~\bibnamefont {Daudet}}, \bibinfo {author}
  {\bibfnamefont {M.}~\bibnamefont {Schuld}}, \bibinfo {author} {\bibfnamefont
  {N.}~\bibnamefont {Tishby}}, \bibinfo {author} {\bibfnamefont
  {L.}~\bibnamefont {Vogt-Maranto}}, \ and\ \bibinfo {author} {\bibfnamefont
  {L.}~\bibnamefont {Zdeborov{\ifmmode\acute{a}\else\'{a}\fi}}},\ }\href
  {\doibase 10.1103/RevModPhys.91.045002} {\bibfield  {journal} {\bibinfo
  {journal} {Rev. Mod. Phys.}\ }\textbf {\bibinfo {volume} {91}},\ \bibinfo
  {pages} {045002} (\bibinfo {year} {2019})}\BibitemShut {NoStop}%
\bibitem [{\citenamefont {Bukov}\ \emph {et~al.}(2018)\citenamefont {Bukov},
  \citenamefont {Day}, \citenamefont {Sels}, \citenamefont {Weinberg},
  \citenamefont {Polkovnikov},\ and\ \citenamefont {Mehta}}]{Mehta2018}%
  \BibitemOpen
  \bibfield  {author} {\bibinfo {author} {\bibfnamefont {M.}~\bibnamefont
  {Bukov}}, \bibinfo {author} {\bibfnamefont {A.~G.~R.}\ \bibnamefont {Day}},
  \bibinfo {author} {\bibfnamefont {D.}~\bibnamefont {Sels}}, \bibinfo {author}
  {\bibfnamefont {P.}~\bibnamefont {Weinberg}}, \bibinfo {author}
  {\bibfnamefont {A.}~\bibnamefont {Polkovnikov}}, \ and\ \bibinfo {author}
  {\bibfnamefont {P.}~\bibnamefont {Mehta}},\ }\href {\doibase
  10.1103/PhysRevX.8.031086} {\bibfield  {journal} {\bibinfo  {journal} {Phys.
  Rev. X}\ }\textbf {\bibinfo {volume} {8}},\ \bibinfo {pages} {031086}
  (\bibinfo {year} {2018})}\BibitemShut {NoStop}%
\bibitem [{\citenamefont {F{\ifmmode\ddot{o}\else\"{o}\fi}sel}\ \emph
  {et~al.}(2018)\citenamefont {F{\ifmmode\ddot{o}\else\"{o}\fi}sel},
  \citenamefont {Tighineanu}, \citenamefont {Weiss},\ and\ \citenamefont
  {Marquardt}}]{Marquardt2018}%
  \BibitemOpen
  \bibfield  {author} {\bibinfo {author} {\bibfnamefont {T.}~\bibnamefont
  {F{\ifmmode\ddot{o}\else\"{o}\fi}sel}}, \bibinfo {author} {\bibfnamefont
  {P.}~\bibnamefont {Tighineanu}}, \bibinfo {author} {\bibfnamefont
  {T.}~\bibnamefont {Weiss}}, \ and\ \bibinfo {author} {\bibfnamefont
  {F.}~\bibnamefont {Marquardt}},\ }\href {\doibase 10.1103/PhysRevX.8.031084}
  {\bibfield  {journal} {\bibinfo  {journal} {Phys. Rev. X}\ }\textbf {\bibinfo
  {volume} {8}},\ \bibinfo {pages} {031084} (\bibinfo {year}
  {2018})}\BibitemShut {NoStop}%
\bibitem [{\citenamefont {Borah}\ \emph {et~al.}(2021)\citenamefont {Borah},
  \citenamefont {Sarma}, \citenamefont {Kewming}, \citenamefont {Milburn},\
  and\ \citenamefont {Twamley}}]{Borah2021}%
  \BibitemOpen
  \bibfield  {author} {\bibinfo {author} {\bibfnamefont {S.}~\bibnamefont
  {Borah}}, \bibinfo {author} {\bibfnamefont {B.}~\bibnamefont {Sarma}},
  \bibinfo {author} {\bibfnamefont {M.}~\bibnamefont {Kewming}}, \bibinfo
  {author} {\bibfnamefont {G.~J.}\ \bibnamefont {Milburn}}, \ and\ \bibinfo
  {author} {\bibfnamefont {J.}~\bibnamefont {Twamley}},\ }\href {\doibase
  10.1103/PhysRevLett.127.190403} {\bibfield  {journal} {\bibinfo  {journal}
  {Phys. Rev. Lett.}\ }\textbf {\bibinfo {volume} {127}},\ \bibinfo {pages}
  {190403} (\bibinfo {year} {2021})}\BibitemShut {NoStop}%
\bibitem [{\citenamefont {Wang}\ \emph {et~al.}(2020)\citenamefont {Wang},
  \citenamefont {Ashida},\ and\ \citenamefont {Ueda}}]{Ueda2020}%
  \BibitemOpen
  \bibfield  {author} {\bibinfo {author} {\bibfnamefont {Z.~T.}\ \bibnamefont
  {Wang}}, \bibinfo {author} {\bibfnamefont {Y.}~\bibnamefont {Ashida}}, \ and\
  \bibinfo {author} {\bibfnamefont {M.}~\bibnamefont {Ueda}},\ }\href {\doibase
  10.1103/PhysRevLett.125.100401} {\bibfield  {journal} {\bibinfo  {journal}
  {Phys. Rev. Lett.}\ }\textbf {\bibinfo {volume} {125}},\ \bibinfo {pages}
  {100401} (\bibinfo {year} {2020})}\BibitemShut {NoStop}%
\bibitem [{\citenamefont {Niu}\ \emph {et~al.}(2019)\citenamefont {Niu},
  \citenamefont {Boixo}, \citenamefont {Smelyanskiy},\ and\ \citenamefont
  {Neven}}]{Niu2019Apr}%
  \BibitemOpen
  \bibfield  {author} {\bibinfo {author} {\bibfnamefont {M.~Y.}\ \bibnamefont
  {Niu}}, \bibinfo {author} {\bibfnamefont {S.}~\bibnamefont {Boixo}}, \bibinfo
  {author} {\bibfnamefont {V.~N.}\ \bibnamefont {Smelyanskiy}}, \ and\ \bibinfo
  {author} {\bibfnamefont {H.}~\bibnamefont {Neven}},\ }\href {\doibase
  10.1038/s41534-019-0141-3} {\bibfield  {journal} {\bibinfo  {journal} {npj
  Quantum Inf.}\ }\textbf {\bibinfo {volume} {5}},\ \bibinfo {pages} {33}
  (\bibinfo {year} {2019})}\BibitemShut {NoStop}%
\bibitem [{\citenamefont {Zhang}\ \emph
  {et~al.}(2019{\natexlab{a}})\citenamefont {Zhang}, \citenamefont {Wei},
  \citenamefont {Asad}, \citenamefont {Yang},\ and\ \citenamefont
  {Wang}}]{Zhang2019Oct}%
  \BibitemOpen
  \bibfield  {author} {\bibinfo {author} {\bibfnamefont {X.-M.}\ \bibnamefont
  {Zhang}}, \bibinfo {author} {\bibfnamefont {Z.}~\bibnamefont {Wei}}, \bibinfo
  {author} {\bibfnamefont {R.}~\bibnamefont {Asad}}, \bibinfo {author}
  {\bibfnamefont {X.-C.}\ \bibnamefont {Yang}}, \ and\ \bibinfo {author}
  {\bibfnamefont {X.}~\bibnamefont {Wang}},\ }\href {\doibase
  10.1038/s41534-019-0201-8} {\bibfield  {journal} {\bibinfo  {journal} {npj
  Quantum Inf.}\ }\textbf {\bibinfo {volume} {5}},\ \bibinfo {pages} {85}
  (\bibinfo {year} {2019}{\natexlab{a}})}\BibitemShut {NoStop}%
\bibitem [{\citenamefont {Xu}\ \emph {et~al.}(2021)\citenamefont {Xu},
  \citenamefont {Wang}, \citenamefont {Yuan},\ and\ \citenamefont
  {Wang}}]{Xu2021Apr}%
  \BibitemOpen
  \bibfield  {author} {\bibinfo {author} {\bibfnamefont {H.}~\bibnamefont
  {Xu}}, \bibinfo {author} {\bibfnamefont {L.}~\bibnamefont {Wang}}, \bibinfo
  {author} {\bibfnamefont {H.}~\bibnamefont {Yuan}}, \ and\ \bibinfo {author}
  {\bibfnamefont {X.}~\bibnamefont {Wang}},\ }\href {\doibase
  10.1103/PhysRevA.103.042615} {\bibfield  {journal} {\bibinfo  {journal}
  {Phys. Rev. A}\ }\textbf {\bibinfo {volume} {103}},\ \bibinfo {pages}
  {042615} (\bibinfo {year} {2021})}\BibitemShut {NoStop}%
\bibitem [{\citenamefont {Zhang}\ \emph
  {et~al.}(2019{\natexlab{b}})\citenamefont {Zhang}, \citenamefont {Wei},
  \citenamefont {Asad}, \citenamefont {Yang},\ and\ \citenamefont
  {Wang}}]{Wang2019a}%
  \BibitemOpen
  \bibfield  {author} {\bibinfo {author} {\bibfnamefont {X.-M.}\ \bibnamefont
  {Zhang}}, \bibinfo {author} {\bibfnamefont {Z.}~\bibnamefont {Wei}}, \bibinfo
  {author} {\bibfnamefont {R.}~\bibnamefont {Asad}}, \bibinfo {author}
  {\bibfnamefont {X.-C.}\ \bibnamefont {Yang}}, \ and\ \bibinfo {author}
  {\bibfnamefont {X.}~\bibnamefont {Wang}},\ }\href {\doibase
  10.1038/s41534-019-0201-8} {\bibfield  {journal} {\bibinfo  {journal} {npj
  Quantum Inf.}\ }\textbf {\bibinfo {volume} {5}},\ \bibinfo {pages} {85}
  (\bibinfo {year} {2019}{\natexlab{b}})}\BibitemShut {NoStop}%
\bibitem [{\citenamefont {Mackeprang}\ \emph {et~al.}(2020)\citenamefont
  {Mackeprang}, \citenamefont {Dasari},\ and\ \citenamefont
  {Wrachtrup}}]{Wrachtrup2020}%
  \BibitemOpen
  \bibfield  {author} {\bibinfo {author} {\bibfnamefont {J.}~\bibnamefont
  {Mackeprang}}, \bibinfo {author} {\bibfnamefont {D.~B.~R.}\ \bibnamefont
  {Dasari}}, \ and\ \bibinfo {author} {\bibfnamefont {J.}~\bibnamefont
  {Wrachtrup}},\ }\href {\doibase 10.1007/s42484-020-00016-8} {\bibfield
  {journal} {\bibinfo  {journal} {Quantum Mach. Intell.}\ }\textbf {\bibinfo
  {volume} {2}},\ \bibinfo {pages} {5} (\bibinfo {year} {2020})}\BibitemShut
  {NoStop}%
\bibitem [{\citenamefont {Haug}\ \emph {et~al.}(2020)\citenamefont {Haug},
  \citenamefont {Mok}, \citenamefont {You}, \citenamefont {Zhang},
  \citenamefont {Png},\ and\ \citenamefont {Kwek}}]{Haug2020}%
  \BibitemOpen
  \bibfield  {author} {\bibinfo {author} {\bibfnamefont {T.}~\bibnamefont
  {Haug}}, \bibinfo {author} {\bibfnamefont {W.-K.}\ \bibnamefont {Mok}},
  \bibinfo {author} {\bibfnamefont {J.-B.}\ \bibnamefont {You}}, \bibinfo
  {author} {\bibfnamefont {W.}~\bibnamefont {Zhang}}, \bibinfo {author}
  {\bibfnamefont {C.~E.}\ \bibnamefont {Png}}, \ and\ \bibinfo {author}
  {\bibfnamefont {L.-C.}\ \bibnamefont {Kwek}},\ }\href {\doibase
  10.1088/2632-2153/abc81f} {\bibfield  {journal} {\bibinfo  {journal} {Mach.
  Learn.: Sci. Technol.}\ }\textbf {\bibinfo {volume} {2}},\ \bibinfo {pages}
  {01LT02} (\bibinfo {year} {2020})}\BibitemShut {NoStop}%
\bibitem [{\citenamefont {Guo}\ \emph {et~al.}(2021)\citenamefont {Guo},
  \citenamefont {Chen}, \citenamefont {Liu}, \citenamefont {Xue}, \citenamefont
  {Chen}, \citenamefont {Cao}, \citenamefont {Mao}, \citenamefont {Tey},\ and\
  \citenamefont {You}}]{Guo2021}%
  \BibitemOpen
  \bibfield  {author} {\bibinfo {author} {\bibfnamefont {S.-F.}\ \bibnamefont
  {Guo}}, \bibinfo {author} {\bibfnamefont {F.}~\bibnamefont {Chen}}, \bibinfo
  {author} {\bibfnamefont {Q.}~\bibnamefont {Liu}}, \bibinfo {author}
  {\bibfnamefont {M.}~\bibnamefont {Xue}}, \bibinfo {author} {\bibfnamefont
  {J.-J.}\ \bibnamefont {Chen}}, \bibinfo {author} {\bibfnamefont {J.-H.}\
  \bibnamefont {Cao}}, \bibinfo {author} {\bibfnamefont {T.-W.}\ \bibnamefont
  {Mao}}, \bibinfo {author} {\bibfnamefont {M.~K.}\ \bibnamefont {Tey}}, \ and\
  \bibinfo {author} {\bibfnamefont {L.}~\bibnamefont {You}},\ }\href {\doibase
  10.1103/PhysRevLett.126.060401} {\bibfield  {journal} {\bibinfo  {journal}
  {Phys. Rev. Lett.}\ }\textbf {\bibinfo {volume} {126}},\ \bibinfo {pages}
  {060401} (\bibinfo {year} {2021})}\BibitemShut {NoStop}%
\bibitem [{\citenamefont {Bilkis}\ \emph {et~al.}(2020)\citenamefont {Bilkis},
  \citenamefont {Rosati}, \citenamefont {Yepes},\ and\ \citenamefont
  {Calsamiglia}}]{Bilkis2020Aug}%
  \BibitemOpen
  \bibfield  {author} {\bibinfo {author} {\bibfnamefont {M.}~\bibnamefont
  {Bilkis}}, \bibinfo {author} {\bibfnamefont {M.}~\bibnamefont {Rosati}},
  \bibinfo {author} {\bibfnamefont {R.~M.}\ \bibnamefont {Yepes}}, \ and\
  \bibinfo {author} {\bibfnamefont {J.}~\bibnamefont {Calsamiglia}},\ }\href
  {\doibase 10.1103/PhysRevResearch.2.033295} {\bibfield  {journal} {\bibinfo
  {journal} {Phys. Rev. Res.}\ }\textbf {\bibinfo {volume} {2}},\ \bibinfo
  {pages} {033295} (\bibinfo {year} {2020})}\BibitemShut {NoStop}%
\bibitem [{\citenamefont {Porotti}\ \emph {et~al.}(2019)\citenamefont
  {Porotti}, \citenamefont {Tamascelli}, \citenamefont {Restelli},\ and\
  \citenamefont {Prati}}]{Prati2019}%
  \BibitemOpen
  \bibfield  {author} {\bibinfo {author} {\bibfnamefont {R.}~\bibnamefont
  {Porotti}}, \bibinfo {author} {\bibfnamefont {D.}~\bibnamefont {Tamascelli}},
  \bibinfo {author} {\bibfnamefont {M.}~\bibnamefont {Restelli}}, \ and\
  \bibinfo {author} {\bibfnamefont {E.}~\bibnamefont {Prati}},\ }\href
  {\doibase 10.1038/s42005-019-0169-x} {\bibfield  {journal} {\bibinfo
  {journal} {{Commun. Phys.}}\ }\textbf {\bibinfo {volume} {{2}}} (\bibinfo
  {year} {{2019}}),\ 10.1038/s42005-019-0169-x}\BibitemShut {NoStop}%
\bibitem [{\citenamefont {Tabuchi}\ \emph {et~al.}(2014)\citenamefont
  {Tabuchi}, \citenamefont {Ishino}, \citenamefont {Ishikawa}, \citenamefont
  {Yamazaki}, \citenamefont {Usami},\ and\ \citenamefont
  {Nakamura}}]{Tabuchi2014Aug}%
  \BibitemOpen
  \bibfield  {author} {\bibinfo {author} {\bibfnamefont {Y.}~\bibnamefont
  {Tabuchi}}, \bibinfo {author} {\bibfnamefont {S.}~\bibnamefont {Ishino}},
  \bibinfo {author} {\bibfnamefont {T.}~\bibnamefont {Ishikawa}}, \bibinfo
  {author} {\bibfnamefont {R.}~\bibnamefont {Yamazaki}}, \bibinfo {author}
  {\bibfnamefont {K.}~\bibnamefont {Usami}}, \ and\ \bibinfo {author}
  {\bibfnamefont {Y.}~\bibnamefont {Nakamura}},\ }\href {\doibase
  10.1103/PhysRevLett.113.083603} {\bibfield  {journal} {\bibinfo  {journal}
  {Phys. Rev. Lett.}\ }\textbf {\bibinfo {volume} {113}},\ \bibinfo {pages}
  {083603} (\bibinfo {year} {2014})}\BibitemShut {NoStop}%
\bibitem [{\citenamefont {Haarnoja}\ \emph {et~al.}(2018)\citenamefont
  {Haarnoja}, \citenamefont {Zhou}, \citenamefont {Abbeel},\ and\ \citenamefont
  {Levine}}]{sac}%
  \BibitemOpen
  \bibfield  {author} {\bibinfo {author} {\bibfnamefont {T.}~\bibnamefont
  {Haarnoja}}, \bibinfo {author} {\bibfnamefont {A.}~\bibnamefont {Zhou}},
  \bibinfo {author} {\bibfnamefont {P.}~\bibnamefont {Abbeel}}, \ and\ \bibinfo
  {author} {\bibfnamefont {S.}~\bibnamefont {Levine}},\ }\href {\doibase
  https://doi.org/10.48550/arXiv.1801.01290} {\enquote {\bibinfo {title} {Soft
  actor-critic: Off-policy maximum entropy deep reinforcement learning with a
  stochastic actor},}\ } (\bibinfo {year} {2018}),\ \Eprint
  {http://arxiv.org/abs/1801.01290} {arXiv:1801.01290 [cs.LG]} \BibitemShut
  {NoStop}%
\bibitem [{\citenamefont {Seberson}\ \emph {et~al.}(2020)\citenamefont
  {Seberson}, \citenamefont {Seberson}, \citenamefont {Ju}, \citenamefont
  {Ahn}, \citenamefont {Bang}, \citenamefont {Li}, \citenamefont {Li},
  \citenamefont {Li}, \citenamefont {Li}, \citenamefont {Robicheaux},\ and\
  \citenamefont {Robicheaux}}]{Seberson2020Dec}%
  \BibitemOpen
  \bibfield  {author} {\bibinfo {author} {\bibfnamefont {T.}~\bibnamefont
  {Seberson}}, \bibinfo {author} {\bibfnamefont {T.}~\bibnamefont {Seberson}},
  \bibinfo {author} {\bibfnamefont {P.}~\bibnamefont {Ju}}, \bibinfo {author}
  {\bibfnamefont {J.}~\bibnamefont {Ahn}}, \bibinfo {author} {\bibfnamefont
  {J.}~\bibnamefont {Bang}}, \bibinfo {author} {\bibfnamefont {T.}~\bibnamefont
  {Li}}, \bibinfo {author} {\bibfnamefont {T.}~\bibnamefont {Li}}, \bibinfo
  {author} {\bibfnamefont {T.}~\bibnamefont {Li}}, \bibinfo {author}
  {\bibfnamefont {T.}~\bibnamefont {Li}}, \bibinfo {author} {\bibfnamefont
  {F.}~\bibnamefont {Robicheaux}}, \ and\ \bibinfo {author} {\bibfnamefont
  {F.}~\bibnamefont {Robicheaux}},\ }\href {\doibase 10.1364/JOSAB.404985}
  {\bibfield  {journal} {\bibinfo  {journal} {J. Opt. Soc. Am. B, JOSAB}\
  }\textbf {\bibinfo {volume} {37}},\ \bibinfo {pages} {3714} (\bibinfo {year}
  {2020})}\BibitemShut {NoStop}%
\bibitem [{\citenamefont {Rusconi}\ \emph {et~al.}(2017)\citenamefont
  {Rusconi}, \citenamefont {P{\"o}chhacker}, \citenamefont {Kustura},
  \citenamefont {Cirac},\ and\ \citenamefont
  {Romero-Isart}}]{Rusconi2017quantum}%
  \BibitemOpen
  \bibfield  {author} {\bibinfo {author} {\bibfnamefont {C.~C.}\ \bibnamefont
  {Rusconi}}, \bibinfo {author} {\bibfnamefont {V.}~\bibnamefont
  {P{\"o}chhacker}}, \bibinfo {author} {\bibfnamefont {K.}~\bibnamefont
  {Kustura}}, \bibinfo {author} {\bibfnamefont {J.~I.}\ \bibnamefont {Cirac}},
  \ and\ \bibinfo {author} {\bibfnamefont {O.}~\bibnamefont {Romero-Isart}},\
  }\href {\doibase 10.1103/PhysRevLett.119.167202} {\bibfield  {journal}
  {\bibinfo  {journal} {Phys. Rev. Lett.}\ }\textbf {\bibinfo {volume} {119}},\
  \bibinfo {pages} {167202} (\bibinfo {year} {2017})}\BibitemShut {NoStop}%
\bibitem [{\citenamefont {Huillery}\ \emph {et~al.}(2020)\citenamefont
  {Huillery}, \citenamefont {Delord}, \citenamefont {Nicolas}, \citenamefont
  {Van Den~Bossche}, \citenamefont {Perdriat},\ and\ \citenamefont
  {H{\ifmmode\acute{e}\else\'{e}\fi}tet}}]{Huillery2020Apr}%
  \BibitemOpen
  \bibfield  {author} {\bibinfo {author} {\bibfnamefont {P.}~\bibnamefont
  {Huillery}}, \bibinfo {author} {\bibfnamefont {T.}~\bibnamefont {Delord}},
  \bibinfo {author} {\bibfnamefont {L.}~\bibnamefont {Nicolas}}, \bibinfo
  {author} {\bibfnamefont {M.}~\bibnamefont {Van Den~Bossche}}, \bibinfo
  {author} {\bibfnamefont {M.}~\bibnamefont {Perdriat}}, \ and\ \bibinfo
  {author} {\bibfnamefont {G.}~\bibnamefont
  {H{\ifmmode\acute{e}\else\'{e}\fi}tet}},\ }\href {\doibase
  10.1103/PhysRevB.101.134415} {\bibfield  {journal} {\bibinfo  {journal}
  {Phys. Rev. B}\ }\textbf {\bibinfo {volume} {101}},\ \bibinfo {pages}
  {134415} (\bibinfo {year} {2020})}\BibitemShut {NoStop}%
\bibitem [{\citenamefont {Forstner}\ \emph {et~al.}(2014)\citenamefont
  {Forstner}, \citenamefont {Sheridan}, \citenamefont {Knittel}, \citenamefont
  {Humphreys}, \citenamefont {Brawley}, \citenamefont {Rubinsztein-Dunlop},\
  and\ \citenamefont {Bowen}}]{Forstner2014UltrasensitiveMagnetometry}%
  \BibitemOpen
  \bibfield  {author} {\bibinfo {author} {\bibfnamefont {S.}~\bibnamefont
  {Forstner}}, \bibinfo {author} {\bibfnamefont {E.}~\bibnamefont {Sheridan}},
  \bibinfo {author} {\bibfnamefont {J.}~\bibnamefont {Knittel}}, \bibinfo
  {author} {\bibfnamefont {C.~L.}\ \bibnamefont {Humphreys}}, \bibinfo {author}
  {\bibfnamefont {G.~A.}\ \bibnamefont {Brawley}}, \bibinfo {author}
  {\bibfnamefont {H.}~\bibnamefont {Rubinsztein-Dunlop}}, \ and\ \bibinfo
  {author} {\bibfnamefont {W.~P.}\ \bibnamefont {Bowen}},\ }\href {\doibase
  10.1002/adma.201401144} {\bibfield  {journal} {\bibinfo  {journal} {Advanced
  Materials}\ }\textbf {\bibinfo {volume} {26}} (\bibinfo {year} {2014}),\
  10.1002/adma.201401144}\BibitemShut {NoStop}%
\bibitem [{\citenamefont {Xia}\ \emph {et~al.}(2015)\citenamefont {Xia},
  \citenamefont {Vanner},\ and\ \citenamefont {Twamley}}]{Xia2015AnQubits}%
  \BibitemOpen
  \bibfield  {author} {\bibinfo {author} {\bibfnamefont {K.}~\bibnamefont
  {Xia}}, \bibinfo {author} {\bibfnamefont {M.~R.}\ \bibnamefont {Vanner}}, \
  and\ \bibinfo {author} {\bibfnamefont {J.}~\bibnamefont {Twamley}},\ }\href
  {\doibase 10.1038/srep05571} {\bibfield  {journal} {\bibinfo  {journal} {Sci.
  Rep.}\ }\textbf {\bibinfo {volume} {4}},\ \bibinfo {pages} {5571} (\bibinfo
  {year} {2015})}\BibitemShut {NoStop}%
\bibitem [{\citenamefont {Yu}\ \emph {et~al.}(2016)\citenamefont {Yu},
  \citenamefont {Janousek}, \citenamefont {Sheridan}, \citenamefont {McAuslan},
  \citenamefont {Rubinsztein-Dunlop}, \citenamefont {Lam}, \citenamefont
  {Zhang},\ and\ \citenamefont {Bowen}}]{Yu2016OptomechanicalResonator}%
  \BibitemOpen
  \bibfield  {author} {\bibinfo {author} {\bibfnamefont {C.}~\bibnamefont
  {Yu}}, \bibinfo {author} {\bibfnamefont {J.}~\bibnamefont {Janousek}},
  \bibinfo {author} {\bibfnamefont {E.}~\bibnamefont {Sheridan}}, \bibinfo
  {author} {\bibfnamefont {D.~L.}\ \bibnamefont {McAuslan}}, \bibinfo {author}
  {\bibfnamefont {H.}~\bibnamefont {Rubinsztein-Dunlop}}, \bibinfo {author}
  {\bibfnamefont {P.~K.}\ \bibnamefont {Lam}}, \bibinfo {author} {\bibfnamefont
  {Y.}~\bibnamefont {Zhang}}, \ and\ \bibinfo {author} {\bibfnamefont {W.~P.}\
  \bibnamefont {Bowen}},\ }\href {\doibase 10.1103/PhysRevApplied.5.044007}
  {\bibfield  {journal} {\bibinfo  {journal} {Phys. Rev. Appl.}\ }\textbf
  {\bibinfo {volume} {5}},\ \bibinfo {pages} {1} (\bibinfo {year}
  {2016})}\BibitemShut {NoStop}%
\bibitem [{\citenamefont {Hoang}\ \emph {et~al.}(2016)\citenamefont {Hoang},
  \citenamefont {Ahn}, \citenamefont {Bang},\ and\ \citenamefont
  {Li}}]{Hoang2016electron}%
  \BibitemOpen
  \bibfield  {author} {\bibinfo {author} {\bibfnamefont {T.~M.}\ \bibnamefont
  {Hoang}}, \bibinfo {author} {\bibfnamefont {J.}~\bibnamefont {Ahn}}, \bibinfo
  {author} {\bibfnamefont {J.}~\bibnamefont {Bang}}, \ and\ \bibinfo {author}
  {\bibfnamefont {T.}~\bibnamefont {Li}},\ }\href {\doibase
  10.1038/ncomms12250} {\bibfield  {journal} {\bibinfo  {journal} {Nat.
  Commun.}\ }\textbf {\bibinfo {volume} {7}},\ \bibinfo {pages} {1} (\bibinfo
  {year} {2016})}\BibitemShut {NoStop}%
\bibitem [{\citenamefont {Delord}\ \emph {et~al.}(2018)\citenamefont {Delord},
  \citenamefont {Huillery}, \citenamefont {Schwab}, \citenamefont {Nicolas},
  \citenamefont {Lecordier},\ and\ \citenamefont
  {H{\'e}tet}}]{Delord2018ramsey}%
  \BibitemOpen
  \bibfield  {author} {\bibinfo {author} {\bibfnamefont {T.}~\bibnamefont
  {Delord}}, \bibinfo {author} {\bibfnamefont {P.}~\bibnamefont {Huillery}},
  \bibinfo {author} {\bibfnamefont {L.}~\bibnamefont {Schwab}}, \bibinfo
  {author} {\bibfnamefont {L.}~\bibnamefont {Nicolas}}, \bibinfo {author}
  {\bibfnamefont {L.}~\bibnamefont {Lecordier}}, \ and\ \bibinfo {author}
  {\bibfnamefont {G.}~\bibnamefont {H{\'e}tet}},\ }\href {\doibase
  10.1103/PhysRevLett.121.053602} {\bibfield  {journal} {\bibinfo  {journal}
  {Phys. Rev. Lett.}\ }\textbf {\bibinfo {volume} {121}},\ \bibinfo {pages}
  {053602} (\bibinfo {year} {2018})}\BibitemShut {NoStop}%
\bibitem [{\citenamefont {Khosla}\ \emph {et~al.}(2017)\citenamefont {Khosla},
  \citenamefont {Brawley}, \citenamefont {Vanner},\ and\ \citenamefont
  {Bowen}}]{khosla2017quantum}%
  \BibitemOpen
  \bibfield  {author} {\bibinfo {author} {\bibfnamefont {K.~E.}\ \bibnamefont
  {Khosla}}, \bibinfo {author} {\bibfnamefont {G.~A.}\ \bibnamefont {Brawley}},
  \bibinfo {author} {\bibfnamefont {M.~R.}\ \bibnamefont {Vanner}}, \ and\
  \bibinfo {author} {\bibfnamefont {W.~P.}\ \bibnamefont {Bowen}},\ }\href
  {\doibase 10.1364/OPTICA.4.001382} {\bibfield  {journal} {\bibinfo  {journal}
  {Optica}\ }\textbf {\bibinfo {volume} {4}},\ \bibinfo {pages} {1382}
  (\bibinfo {year} {2017})}\BibitemShut {NoStop}%
\bibitem [{\citenamefont {Achiam}(2021)}]{SpinningUp2021}%
  \BibitemOpen
  \bibfield  {author} {\bibinfo {author} {\bibfnamefont {J.}~\bibnamefont
  {Achiam}},\ }\href {https://github.com/openai/spinningup} {\enquote {\bibinfo
  {title} {{Spinning Up in Deep Reinforcement Learning}},}\ } (\bibinfo {year}
  {2021}),\ \bibinfo {note} {[Online; accessed 15. Jul. 2021]}\BibitemShut
  {NoStop}%
\bibitem [{\citenamefont {Brockman}\ \emph {et~al.}(2016)\citenamefont
  {Brockman}, \citenamefont {Cheung}, \citenamefont {Pettersson}, \citenamefont
  {Schneider}, \citenamefont {Schulman}, \citenamefont {Tang},\ and\
  \citenamefont {Zaremba}}]{OpenaiGym}%
  \BibitemOpen
  \bibfield  {author} {\bibinfo {author} {\bibfnamefont {G.}~\bibnamefont
  {Brockman}}, \bibinfo {author} {\bibfnamefont {V.}~\bibnamefont {Cheung}},
  \bibinfo {author} {\bibfnamefont {L.}~\bibnamefont {Pettersson}}, \bibinfo
  {author} {\bibfnamefont {J.}~\bibnamefont {Schneider}}, \bibinfo {author}
  {\bibfnamefont {J.}~\bibnamefont {Schulman}}, \bibinfo {author}
  {\bibfnamefont {J.}~\bibnamefont {Tang}}, \ and\ \bibinfo {author}
  {\bibfnamefont {W.}~\bibnamefont {Zaremba}},\ }\href@noop {} {\enquote
  {\bibinfo {title} {Openai gym},}\ } (\bibinfo {year} {2016}),\ \Eprint
  {http://arxiv.org/abs/1606.01540} {arXiv:1606.01540 [cs.LG]} \BibitemShut
  {NoStop}%
\bibitem [{\citenamefont {Raffin}\ \emph {et~al.}(2019)\citenamefont {Raffin},
  \citenamefont {Hill}, \citenamefont {Ernestus}, \citenamefont {Gleave},
  \citenamefont {Kanervisto},\ and\ \citenamefont
  {Dormann}}]{stable-baselines3}%
  \BibitemOpen
  \bibfield  {author} {\bibinfo {author} {\bibfnamefont {A.}~\bibnamefont
  {Raffin}}, \bibinfo {author} {\bibfnamefont {A.}~\bibnamefont {Hill}},
  \bibinfo {author} {\bibfnamefont {M.}~\bibnamefont {Ernestus}}, \bibinfo
  {author} {\bibfnamefont {A.}~\bibnamefont {Gleave}}, \bibinfo {author}
  {\bibfnamefont {A.}~\bibnamefont {Kanervisto}}, \ and\ \bibinfo {author}
  {\bibfnamefont {N.}~\bibnamefont {Dormann}},\ }\href@noop {} {\enquote
  {\bibinfo {title} {Stable baselines3},}\ }\bibinfo {howpublished}
  {\url{https://github.com/DLR-RM/stable-baselines3}} (\bibinfo {year}
  {2019})\BibitemShut {NoStop}%
\bibitem [{\citenamefont {Paszke}\ \emph {et~al.}(2019)\citenamefont {Paszke},
  \citenamefont {Gross}, \citenamefont {Massa}, \citenamefont {Lerer},
  \citenamefont {Bradbury}, \citenamefont {Chanan}, \citenamefont {Killeen},
  \citenamefont {Lin}, \citenamefont {Gimelshein}, \citenamefont {Antiga},
  \citenamefont {Desmaison}, \citenamefont {Kopf}, \citenamefont {Yang},
  \citenamefont {DeVito}, \citenamefont {Raison}, \citenamefont {Tejani},
  \citenamefont {Chilamkurthy}, \citenamefont {Steiner}, \citenamefont {Fang},
  \citenamefont {Bai},\ and\ \citenamefont {Chintala}}]{pytorch}%
  \BibitemOpen
  \bibfield  {author} {\bibinfo {author} {\bibfnamefont {A.}~\bibnamefont
  {Paszke}}, \bibinfo {author} {\bibfnamefont {S.}~\bibnamefont {Gross}},
  \bibinfo {author} {\bibfnamefont {F.}~\bibnamefont {Massa}}, \bibinfo
  {author} {\bibfnamefont {A.}~\bibnamefont {Lerer}}, \bibinfo {author}
  {\bibfnamefont {J.}~\bibnamefont {Bradbury}}, \bibinfo {author}
  {\bibfnamefont {G.}~\bibnamefont {Chanan}}, \bibinfo {author} {\bibfnamefont
  {T.}~\bibnamefont {Killeen}}, \bibinfo {author} {\bibfnamefont
  {Z.}~\bibnamefont {Lin}}, \bibinfo {author} {\bibfnamefont {N.}~\bibnamefont
  {Gimelshein}}, \bibinfo {author} {\bibfnamefont {L.}~\bibnamefont {Antiga}},
  \bibinfo {author} {\bibfnamefont {A.}~\bibnamefont {Desmaison}}, \bibinfo
  {author} {\bibfnamefont {A.}~\bibnamefont {Kopf}}, \bibinfo {author}
  {\bibfnamefont {E.}~\bibnamefont {Yang}}, \bibinfo {author} {\bibfnamefont
  {Z.}~\bibnamefont {DeVito}}, \bibinfo {author} {\bibfnamefont
  {M.}~\bibnamefont {Raison}}, \bibinfo {author} {\bibfnamefont
  {A.}~\bibnamefont {Tejani}}, \bibinfo {author} {\bibfnamefont
  {S.}~\bibnamefont {Chilamkurthy}}, \bibinfo {author} {\bibfnamefont
  {B.}~\bibnamefont {Steiner}}, \bibinfo {author} {\bibfnamefont
  {L.}~\bibnamefont {Fang}}, \bibinfo {author} {\bibfnamefont {J.}~\bibnamefont
  {Bai}}, \ and\ \bibinfo {author} {\bibfnamefont {S.}~\bibnamefont
  {Chintala}},\ }in\ \href
  {http://papers.neurips.cc/paper/9015-pytorch-an-imperative-style-high-performance-deep-learning-library.pdf}
  {\emph {\bibinfo {booktitle} {Advances in Neural Information Processing
  Systems 32}}},\ \bibinfo {editor} {edited by\ \bibinfo {editor}
  {\bibfnamefont {H.}~\bibnamefont {Wallach}}, \bibinfo {editor} {\bibfnamefont
  {H.}~\bibnamefont {Larochelle}}, \bibinfo {editor} {\bibfnamefont
  {A.}~\bibnamefont {Beygelzimer}}, \bibinfo {editor} {\bibfnamefont
  {F.}~\bibnamefont {d\textquotesingle Alch\'{e}-Buc}}, \bibinfo {editor}
  {\bibfnamefont {E.}~\bibnamefont {Fox}}, \ and\ \bibinfo {editor}
  {\bibfnamefont {R.}~\bibnamefont {Garnett}}}\ (\bibinfo  {publisher} {Curran
  Associates, Inc.},\ \bibinfo {year} {2019})\ pp.\ \bibinfo {pages}
  {8024--8035}\BibitemShut {NoStop}%
\bibitem [{\citenamefont {Fletcher}\ and\ \citenamefont
  {Bell}(1959)}]{Fletcher1959ferrimagnetic}%
  \BibitemOpen
  \bibfield  {author} {\bibinfo {author} {\bibfnamefont {P.}~\bibnamefont
  {Fletcher}}\ and\ \bibinfo {author} {\bibfnamefont {R.}~\bibnamefont
  {Bell}},\ }\href {\doibase 10.1063/1.1735216} {\bibfield  {journal} {\bibinfo
   {journal} {J. Appl. Phys.}\ }\textbf {\bibinfo {volume} {30}},\ \bibinfo
  {pages} {687} (\bibinfo {year} {1959})}\BibitemShut {NoStop}%
\bibitem [{\citenamefont {Wang}\ and\ \citenamefont
  {Clerk}(2012)}]{Wang2012using}%
  \BibitemOpen
  \bibfield  {author} {\bibinfo {author} {\bibfnamefont {Y.-D.}\ \bibnamefont
  {Wang}}\ and\ \bibinfo {author} {\bibfnamefont {A.~A.}\ \bibnamefont
  {Clerk}},\ }\href {\doibase 10.1103/PhysRevLett.108.153603} {\bibfield
  {journal} {\bibinfo  {journal} {Phys. Rev. Lett.}\ }\textbf {\bibinfo
  {volume} {108}},\ \bibinfo {pages} {153603} (\bibinfo {year}
  {2012})}\BibitemShut {NoStop}%
\bibitem [{\citenamefont {Bergmann}\ \emph {et~al.}(1998)\citenamefont
  {Bergmann}, \citenamefont {Theuer},\ and\ \citenamefont
  {Shore}}]{Bergmann1998Jul}%
  \BibitemOpen
  \bibfield  {author} {\bibinfo {author} {\bibfnamefont {K.}~\bibnamefont
  {Bergmann}}, \bibinfo {author} {\bibfnamefont {H.}~\bibnamefont {Theuer}}, \
  and\ \bibinfo {author} {\bibfnamefont {B.~W.}\ \bibnamefont {Shore}},\ }\href
  {\doibase 10.1103/RevModPhys.70.1003} {\bibfield  {journal} {\bibinfo
  {journal} {Rev. Mod. Phys.}\ }\textbf {\bibinfo {volume} {70}},\ \bibinfo
  {pages} {1003} (\bibinfo {year} {1998})}\BibitemShut {NoStop}%
\end{thebibliography}
%%%%%%%


\end{document}